\documentclass[prd,aps,nofootinbib,showpacs,preprintnumbers,amsmath,amssymb,floatfix]{revtex4}

\usepackage{times}
\usepackage{hyperref}

\usepackage{graphicx}% include figure files
\usepackage{dcolumn}% Align table columns on decimal point
\usepackage{bm}% bold math
\usepackage{slashed}

\begin{document}

\title{Precision radiative corrections to the muon polarization in the semileptonic decay of a charged kaon
}

\author{
M.~J.\ S\'anchez-Glez, A.\ Mart{\'\i}nez
}
\affiliation{
Departamento de F{\'\i}sica, Escuela Superior de F\'{\i}sica y Matem\'aticas del Instituto Polit\'ecnico Nacional, Apartado Postal 75-702, Ciudad de M\'exico 07738, Mexico
}

\author{
C.\ Ju\'arez-Le\'on
}
\affiliation{
Departamento de Formaci\'on B\'asica, Escuela Superior de C\'omputo del Instituto Polit\'ecnico Nacional, Apartado Postal 75-702, Ciudad de M\'exico 07738, Mexico
}

\author{
M.\ Neri
}
\affiliation{
Departamento de F{\'\i}sica, Escuela Superior de F\'{\i}sica y Matem\'aticas del Instituto Polit\'ecnico Nacional, Apartado Postal 75-702, Ciudad de M\'exico 07738, Mexico
}

\author{
J.\ J.\ Torres
}
\affiliation{
Departamento de Posgrado, Escuela Superior de C\'omputo del Instituto Polit\'ecnico Nacional, Apartado Postal 75-702, Ciudad de M\'exico 07738, Mexico
}

\author{
Rub\'en Flores-Mendieta
}
\affiliation{
Instituto de F{\'\i}sica, Universidad Aut\'onoma de San Luis Potos{\'\i}, \'Alvaro Obreg\'on 64, Zona Centro, San Luis Potos{\'\i}, S.L.P.\ 78000, Mexico
}

\date{\today}

\begin{abstract}
An expression for the Dalitz plot of the semileptonic decay of a charged kaon, including radiative corrections to order $\mathcal{O}[(\alpha/\pi)(q/M_1)]$, where $q$ is the four-momentum transfer and $M_1$ is the mass of the decaying kaon, is obtained. Contributions of both the three- and four-body regions are accounted for. Besides, the emitted muon is considered to be polarized so the analysis is also focused on evaluating the radiative corrections to the longitudinal, transverse, and normal polarization components of the muon. The final formulas, with the triple integration of the bremsstrahlung photon variables ready to be performed numerically, are general enough to be used in model-independent experimental analyses whether or not the real photon is discriminated. With the numerical values of the weak form factors and slope parameters of the process, the radiative corrections to the components of the muon polarization are found to be very small compared to their respective uncorrected values.
\end{abstract}


\maketitle

\section{Introduction}

The symmetry transformation CPT is the product of the three symmetry transformations C, P, and T, which interchange particles and antiparticles, $\mathbf{x}$ and $-\mathbf{x}$, and $t$ and $-t$, respectively. As long as CPT invariance holds, both CP and T violations are accounted for by complex relative phases in the effective coupling constants. For a CP or T asymmetry to become observable at least two amplitudes containing different coupling constants are required \cite{bigi}.

Kaon physics has played an important role in all the stages of the Standard Model (SM), including its very construction and its tests through CP violation in $K_L \to \pi\pi$ decays, to name but a few. Direct CP violation can manifest itself through a difference in the normalized decay amplitudes for $K_L\to \pi^+\pi^-$ with respect to $K_L\to \pi^0\pi^0$; significant experimental effort has been carried out to understand the mechanism of CP violation \cite{cronin,na31,e731,ktev}. T violation, on the other hand, was first suggested by Sakurai to be searched in the transverse muon polarization $P_T$ in the decay $K^+\to \pi^0\mu^+\nu$ ($K_{\mu 3}^+$) \cite{saku}; $P_T$, a T-odd observable, is the polarization component normal to the decay plane and is defined by the correlation of the $\pi^0$ and $\mu^+$ momentum vectors and the $\mu^+$ spin vector. A nonzero value of $P_T$, at the level of $\mathcal{O}(10^{-3})$ down to $\mathcal{O}(10^{-5})$, would indicate a clear evidence for violation of time reversal invariance since the spurious effects from final state interactions are negligible, so the SM expectations for $P_T$ are around $\mathcal{O}(10^{-7})$ \cite{bigi}. Various experimental setups have performed careful analyses to determine $P_T$ \cite{morse,kek}, which have found T symmetry within one standard deviation. Particularly, the upper limit $|P_T| < 0.0050$ at 90\% confidence level has been obtained by the KEK-E246 Collaboration \cite{kek}, which has set constraints on the parameters of different theoretical models beyond the SM.

On the other hand, the observable quantities in $K_{\ell 3}$ decays depend on the $V_{us}$ matrix element and on two form factors $f_\pm(q^2)$, where $q$ is the momentum transfer. A precise determination of the latter provides crucial information about the low-energy dynamics of the strong interactions. There are various theoretical analyses of the form factors; some of them are based on the quark model (or variations of it) but a more modern determination has been carried out in the context of quiral perturbation theory. Indeed, the seminal papers by Cirigliano and collaborators \cite{ciri,ciri2} presented detailed computations to leading nontrivial order in the chiral effective field theory and have shed light into the subject.

On the experimental bent, precise measurements have been made so far on $K_{\ell 3}$ decays \cite{part}. The statistical errors are small. To achieve precise values for $V_{us}$ and the form factors, the reduction of the various systematical errors must be seriously considered. There are two types of systematic errors, one of which has to do with the different shortcomings of the experimental devices, and the other of which is of theoretical nature. The main sources of theoretical uncertainties are
\begin{enumerate}
\item[(a)] Assumptions for the form factors.
\item[(b)] Momentum transfer dependence of the form factors.
\item[(c)] Radiative corrections (RC) to the integrated observables.
\end{enumerate}

Case (a) listed above can be analyzed using the approach discussed in Refs.~\cite{ciri,ciri2}, or any other approach that evaluates flavor $SU(3)$ breaking effects in the form factors, consistent with the Ademollo-Gatto theorem. Case (b) can be dealt with the methods suggested in Ref.~\cite{part}, which are summarized in the following section. And case (c), which encompasses the precise and reliable calculation of RC to various measurable quantities relevant for experimental analyses, is a rather involved theoretical task. The subject was formerly dealt with in Refs.~\cite{gins66,gins69,gins70}, also in Refs.~\cite{ciri,ciri2}, and relatively more recently in Refs.~\cite{juarez11,juarez12,juarez15,neri16}, specializing the analysis to the Dalitz plot (DP) of the dacay. Henceforth, the DP will be referred to as the differential decay rate of $K_{\ell 3}$ decays, expressed in terms of the pion and charged lepton energies

In particular, the method discussed in Refs.~\cite{juarez11,juarez12,juarez15,neri16} to compute RC to the DP of $K_{\ell 3}$ decays to order $\mathcal{O}[(\alpha/\pi)(q/M_1)]$, where $M_1$ is the mass of the kaon, lead to expressions that are suitable for model-independent analyses. The model dependence of the virtual RC was handled by implementing the method of Sirlin \cite{sirlin}, originally introduced in the RC of neutron $\beta$ decay, while the model dependence of the bremsstrahlung RC was under control by virtue of the Low theorem \cite{low,chew}.

Dealing with bremsstrahlung RC was not an easy matter: It is a four-body decay whose DP covers completely the DP of the three-body decay. The latter is referred to as the three-body region (TBR) and the non-overlap of the former and the latter as the four-body region (FBR). Among other properties, the final expressions obtained did not contain infrared divergences, did not depend on an ultraviolet cutoff, and did not contain any model dependence of RC. The model dependence was absorbed into the already existing form factors, which, as a side remark, were not compromised to be fixed at predetermined values.

Once the RC to the DP are available, the next natural step is to extend the approach to cover some other observables. One of them is precisely the muon polarization vector. From the theoretical point of view, the evaluation of the muon polarization should be a rather straightforward problem because the observable effects of spin polarization can be easily implemented at the level of the transition amplitude. The muon is one-hundred polarized at each point of the DP, and the components of the polarization vector depend on the value of the parameter $\xi(q^2)\equiv f_-(q^2)/f_+(q^2)$ at that point. Cabibbo and Maksymowicz pointed out that a precise measurement of the muon polarization would yield a sensitive method to determine $\xi(q^2)$ \cite{cym}. Formally, RC can decrease the values of the components of the muon polarization at each point of the DP. Therefore, it is required that, in addition to high-statistics experiments, theoretical expressions as general and accurate as possible to evaluate those RC be available.

The analyses of RC to order $\mathcal{O}(\alpha)$ to the muon polarization have been performed in Refs.~\cite{gins71,gins73} for $K_{\mu 3}^+$ and $K_{\mu 3}^0$ decays, respectively. The approach is based on a phenomenological weak $K$-$\pi$ vertex and the resultant expressions depend on a logarithmic cutoff. 

In the present paper, the issue of RC to order $\mathcal{O}[(\alpha/\pi)(q/M_1)]$ to the muon polarization in the semileptonic decay of a positively charged kaon will be evaluated, taking into account contributions of both the TBR and the FBR. A word of caution is in order here. It should be stressed that the issue of T violation is not, by far, the subject of the present paper. Instead, the main goal is to produce accurate theoretical expressions to evaluate the RC to three components of the muon polarization using the available data on the form factors.

This paper is organized in the following way. In Sec.~\ref{sec:intro} an overview of kaon semileptonic decays is provided in order to set notation and conventions. In Sec.~\ref{sec:vrc} the analysis of virtual RC is discussed, putting special attention on the spin-dependent part because its counterpart, the spin-independent one, has been discussed in Ref.~\cite{juarez11}. In Sec.~\ref{sec:brc} the bremsstrahlung RC are discussed for both the TBR and the FBR of the DP. Several triple integrals---mainly over the photon variables---emerge. Any attempts to analytically evaluate those integrals for all three components of the muon polarization are fruitless because several elliptic integrals appear. Except for the longitudinal polarization, the integrals are computed numerically. The complete DP is presented in Sec.~\ref{sec:DPComplete}. In Sec.~\ref{sec:mpol} the muon polarization vector, including RC, is explicitly given and its components are numerically evaluated at several points of the allowed kinematical region. A discussion of the findings is presented in Sec~\ref{sec:closing}. The paper is complemented with an appendix where the results of two new integrals appearing in the longitudinal polarization are shown.

\section{\label{sec:intro}An overview of kaon semileptonic decays}

The present analysis builds on earlier studies about RC in kaon semileptonic decays \cite{juarez11,juarez12,juarez15,neri16}, so only an overview of the subject is provided here in order to introduce the notation and conventions.

For definiteness, the semileptonic decay of a positively charged kaon, hereafter referred to as $K_{\ell 3}^+$, is represented by
\begin{equation}
K^+(p_1) \to \pi^0(p_2) + \ell^+(l) + \nu_\ell (p_\nu), \label{eq:kl3p}
\end{equation}
where the four-momenta and masses of the $K^+$, $\pi^0$, $\ell^+$, and $\nu_\ell$ are denoted by $p_1=(E_1,{\mathbf p}_1)$, $p_2=(E_2,{\mathbf p}_2)$, $l=(E,{\mathbf l})$, and $p_\nu=(E_\nu^0,{\mathbf p}_\nu^0)$, and by $M_1$, $M_2$, $m$, and $m_\nu$, respectively. In particular, the decay mode with $\ell=\mu$ is the case study here. The reference system used is the rest frame of $K^+$, to which all noncovariant expressions in this paper will refer. In this context, quantities like $p_2$, $l$, or $p_\nu$ will also stand for the magnitudes of the corresponding three-momenta, unless explicitly noticed otherwise. Additionally, the direction of a generic three-vector ${\mathbf p}$ will be denoted by a unit vector $\hat {\mathbf p}$.

The uncorrected transition amplitude (\textit{i.e.}\ the amplitude without RC) for process~(\ref{eq:kl3p}) is given, in the context of the $V-A$ theory, by
\begin{eqnarray}
\mathsf{M}_0 = C_K \frac{G_F}{\sqrt 2} V_{us}^* \left[ f_+(q^2)(p_1+p_2)_\alpha +f_-(q^2)(p_1-p_2)_\alpha \right] \left[\overline{u}_\nu(p_\nu) O_\alpha v_\ell(l)\right], \label{eq:M0}
\end{eqnarray}
where $C_K=1/\sqrt{2}$ is a Clebsch-Gordan coefficient, $G_F$ is the Fermi constant as extracted from muon decay, $V_{us}$ is the relevant element of the Cabibbo-Kobayashi-Maskawa (CKM) matrix, and $f_\pm(q^2)$ are the usual form factors which depend on the four-momentum transfer $q\equiv p_1-p_2$. Here, $v_\ell$ and $u_\nu$ are the Dirac spinors of the corresponding particles, $O_\alpha \equiv \gamma_\alpha(1+\gamma_5)$, and the metric and $\gamma$-matrix convention adopted are specified in Ref.~\cite{juarez11}.

A common practice advocated in the analysis of $K_{\mu 3}$ decays is to assume a linear dependence of $f_\pm$ on $q^2$, namely,
\begin{equation}
f_\pm(q^2) = f_\pm(0) \left[ 1 + \lambda_\pm \frac{q^2}{M_2^2} \right],
\end{equation}
and most data are properly described with a constant $f_-$ \cite{part}.

An equivalent parametrization of the form factors introduces the ratio
\begin{equation}
\xi(q^2) \equiv \frac{f_-(q^2)}{f_+(q^2)},
\end{equation}
so the relevant parameters are $\lambda_+$ and $\xi(0)$.

Recent analyses, however, use the form factors $f_+$ and $f_0$ instead, which are respectively related to vector and scalar exchange to the lepton pair. The relation among them is \cite{part}
\begin{equation}
f_0(q^2) = f_+(q^2) + \frac{q^2}{M_1^2-M_2^2} f_-(q^2).
\end{equation}

For a $f_+$ linear in $q^2$ and $f_-$ constant, $f_0(q^2)$ can be expressed as
\begin{equation}
f_0(q^2) = f_0(0) \left[ 1 + \lambda_0 \frac{q^2}{M_2^2} \right],
\end{equation}
and, under the same assumptions, the parameter $\xi(q^2)$ can be rewritten as
\begin{eqnarray}
\xi(q^2) & = & \frac{M_1^2-M_2^2}{M_2^2} \frac{\lambda_0 - \lambda_+}{1 + \lambda_+ \displaystyle \frac{q^2}{M_2^2}} \nonumber \\
& = & \xi(0) \left[1 + \lambda_+ \frac{q^2}{M_2^2} \right]^{-1}. \label{eq:xiq2}
\end{eqnarray}
Presumably, the $(\lambda_+,\lambda_0)$ correlations tend to be less strong than the $(\lambda_+,\xi(0))$ correlations \cite{part}. For numerical purposes only, henceforth the values $\mathrm{Re} \, \xi(0) = -0.126 \pm 0.016$ [obtained using Eq.~(\ref{eq:xiq2}) assuming linear energy dependence of $f_+(0)$ and $\mu$-$e$ universality] and $\mathrm{Im} \, \xi(0) = -0.006 \pm 0.008$ (suggested in Ref.~\cite{part}) will be used; as for the dependence on $q^2$, the former is assumed to follow expression (\ref{eq:xiq2}) whereas the latter is assumed to be constant.

In order to analyze the muon polarization, a more convenient form of the transition amplitude can thus be written as
\begin{equation}
\mathsf{M}_0 = C_K \frac{G_F}{\sqrt 2} V_{us}^* f_+(q^2) \left\{ 2{p_1}_\alpha-[1-\xi(q^2)] q_\alpha \right\} [\overline{u}_\nu(p_\nu) O_\alpha v_\ell(l)]. \label{eq:M0a}
\end{equation}

The polarization of the emitted muon can be accounted for by introducing the spin projection operator
\begin{equation}
\Sigma(s) = \frac{1-\gamma_5 \slashed{s}}{2}, \label{eq:sproj}
\end{equation}
where $s\cdot s =s_0^2 - \mathbf{s}\cdot \mathbf{s}=-1$ and $s\cdot l=0$. In the rest frame of the muon, $s$ reduces to a purely spatial unit vector $\hat{\mathbf{s}}_R$ which gives the spin direction. The observable effects of spin polarization can then be analyzed through the replacement
\begin{equation}
v_\ell(l) \to \Sigma(s) v_\ell(l), \label{eq:speff}
\end{equation}
in the corresponding spinor of the muon in the transition amplitude $\mathsf{M}_0$.

The usual procedure to obtain the differential decay rate requires the calculation of the decay amplitude squared, performing a summation over the spins in the final state. In the present case, this procedure yields,
\begin{equation}
\sum_{\mathrm{spins}} |\mathsf{M}_0|^2 = \frac12 \sum_{\mathrm{spins}} |\mathsf{M}_0^\prime|^2 - \frac12 \sum_{\mathrm{spins}} |\mathsf{M}_0^{(s)}|^2, \label{eq:m0sq}
\end{equation}
where the first and second summands in Eq.~(\ref{eq:m0sq}) comprise the spin-independent and spin-dependent contributions to $\sum |\mathsf{M}_0|^2$, respectively.

The uncorrected differential decay rate for process (\ref{eq:kl3p}), represented here by $d\Gamma_0$, can be given by \cite{juarez11}
\begin{equation}
d\Gamma_0 = \frac{1}{2M_1} \frac{d^3p_2}{2E_2(2\pi)^3} \frac{m}{E} \frac{d^3l}{(2\pi)^3} \frac{m_\nu}{E_\nu^0} \frac{d^3p_\nu^0}{(2\pi)^3} (2\pi)^4 \delta^4(p_1-p_2-l-p_\nu)\sum_{\textrm{spins}}|\mathsf{M}_0|^2.
\end{equation}
$d\Gamma_0$ is most conveniently evaluated in the kaon rest system, leaving the energies of the muon and pion, $E$ and $E_2$, as independent variables, which yields the so-called DP. To this end, the integral over the three-momentum of the neutrino can be performed straightforwardly. For the integrals over the angular variables of the muon and pion, the coordinate axes can be oriented in such a way that $\ell^+$ be emitted along the $+z$ axis and $\pi^0$ be emitted in the first or fourth quadrant of the $(x,z)$ plane. Thus, the only nontrivial angular integration left is over the polar angle of $\pi^0$, $\theta_2$, namely
\begin{equation}
d\Gamma_0 = \frac{1}{(2\pi)^3}\frac{mm_\nu}{2M_1}dEdE_2\int_{-1}^1dy\delta(y-y_0) \sum_{\textrm{spins}}|\mathsf{M}_0|^2,
\end{equation}
where $y=\cos\theta_2$ and
\begin{equation}
y_0 = \frac{{E_\nu^0}^2-p_2^2-l^2}{2p_2l} \label{eq:y0}
\end{equation}
is the cosine of the angle between $\mathbf{p}_2$ and $\mathbf{l}$. By energy conservation, $E_\nu^0 = M_1 - E_2 -E$.

\subsection{Muon polarization}

At this stage, the corresponding spin-independent and spin-dependent contributions to the total decay rate, $d\Gamma_0^\prime$ and $d\Gamma_0^{(s)}$, can readily be obtained. The total decay rate is then expressed as
\begin{equation}
d\Gamma_0(K_{\mu 3}^+) = \frac12 d\Gamma_0^\prime + \frac12 d\Gamma_0^{(s)}.
\end{equation}

The muon spin $s$ in the kaon rest frame is related to its vector $\hat{\mathbf{s}}_R$ in the muon rest frame by
\begin{equation}
s_0 = \frac{1}{m} \hat{\mathbf{s}}_R \cdot \mathbf{l}, \qquad \mathbf{s}_{\|} = \frac{E}{m}(\hat{\mathbf{s}}_R \cdot \hat{\mathbf{l}}) \hat{\mathbf{l}}, \qquad \mathbf{s}_\bot = \hat{\mathbf{s}}_R-(\hat{\mathbf{s}}_R \cdot \hat{\mathbf{l}}) \hat{\mathbf{l}},
\end{equation}
so that, if $u=(u_0,\mathbf{u})$ is an arbitrary four-vector, then
\begin{equation}
s\cdot u = \hat{\mathbf{s}}_R\cdot \left[\frac{\mathbf{l}}{m} \left(u_0 - \frac{\mathbf{u} \cdot \mathbf{l}}{E+m}\right) - \mathbf{u} \right].
\end{equation}

For definiteness, let the decay plane be spanned by the vectors $\mathbf{l}$ and $\mathbf{p}_2$; three orthogonal components of the uncorrected muon polarization vector $\mathbf{P}_0$ can thus be defined, namely, the longitudinal $P_{L0}$, transverse $P_{T0}$, and normal $P_{N0}$, as\footnote{For the ease of notation and unless explicitly noticed otherwise, $\mathbf{P}_0$ and its three components are understood to depend on $E$ and $E_2$.}
\begin{subequations}
\label{eq:Pcomp}
\begin{eqnarray}
&  & P_{L0} = \mathbf{P}_0 \cdot \hat{\bm{\epsilon}}_L, \\
&  & P_{T0} = \mathbf{P}_0 \cdot \hat{\bm{\epsilon}}_T, \\
&  & P_{N0} = \mathbf{P}_0 \cdot \hat{\bm{\epsilon}}_N,
\end{eqnarray}
\end{subequations}
where the subscript 0 denotes an uncorrected quantity. On the other hand, the vectors $\hat{\bm{\epsilon}}_L$, $\hat{\bm{\epsilon}}_T$, and $\hat{\bm{\epsilon}}_N$ form an orthonormal basis and are defined as
\begin{subequations}
\label{eq:basis}
\begin{eqnarray}
\hat{\bm{\epsilon}}_L & \equiv & \frac{\mathbf{l}}{\left|\mathbf{l}\right|}, \\
\hat{\bm{\epsilon}}_T & \equiv & \frac{\mathbf{p}_2 \times \mathbf{l}}{\mathbf{\left|p_{2}\times\mathbf{l}\right|}}, \\
\hat{\bm{\epsilon}}_N & \equiv & \hat{\bm{\epsilon}}_L \times \hat{\bm{\epsilon}}_T.
\end{eqnarray}
\end{subequations}
Here $\hat{\bm{\epsilon}}_L$ is parallel to the muon momentum, $\hat{\bm{\epsilon}}_T$ is perpendicular to the decay plane, and $\hat{\bm{\epsilon}}_N$ is normal to both $\hat{\bm{\epsilon}}_L$ and $\hat{\bm{\epsilon}}_T$ in the decay plane.

Based on the above premises, the spin-independent contribution to the uncorrected differential decay rate reads
\begin{equation}
d\Gamma_0^\prime = a_0^\prime d\Omega^\prime , \label{eq:Gama0sin}
\end{equation}
where $a_0^\prime$ depends on the energies $E$ and $E_2$, the masses of the particles involved in the process and quadratically on the parameter $\xi(q^2)$. Its explicit form is
\begin{equation}
a_0^\prime = (2M_1E-m^2) E_\nu^0  - \left[ M_1^2 - \frac{m^2}{4} \right] \frac{q^2-m^2}{2M_1} + m^2\left[ E_\nu^0 -\frac{q^2-m^2}{4M_1} \right]\mathrm{Re} \, \xi(q^2) + \frac{m^2(q^2-m^2)}{8M_1} \left|\xi(q^2)\right|^2, \label{eq:Amp0sin}
\end{equation}
where
\begin{equation}
d\Omega^\prime = \frac{C_K^2 G_F^2 |V_{us}|^2 dEdE_2}{4\pi^3} |f_+(q^2)|^2. \label{eq:domega}
\end{equation}

Similarly, the spin-dependent contribution is
\begin{equation}
d\Gamma_0^{(s)} = \hat{\mathbf{s}}_R \cdot \mathbf{a}_0^{(s)} d\Omega^\prime,
\end{equation}
where $\mathbf{a}_0^{(s)}$ is a vector which also depends on the energies $E$ and $E_2$, the masses of the particles involved in the process and quadratically on the parameter $\xi(q^2)$. In terms of the orthonormal basis, it reads,
\begin{equation}
\mathbf{a}_0^{(s)} = \Lambda_{L0} \hat{\bm{\epsilon}}_L + \Lambda_{T0} \hat{\bm{\epsilon}}_T + \Lambda_{N0} \hat{\bm{\epsilon}}_N, \label{eq:a0s}
\end{equation}
where the different $\Lambda_{X0}$ functions ($X=L,T,N$) are defined as
\begin{equation}
\Lambda_{L0} = M_1lE_\nu^0 + (l + p_2y_0)(m^2-M_1E) + C_{L0} - \left[ m^2(l + p_2y_0) + 2 C_{L0} \right] \mathrm{Re} \, \xi(q^2) + C_{L0}\ |\xi(q^2)|^2,
\end{equation}
\begin{equation}
\Lambda_{T0} = m p_2l \sqrt{1-y_0^2} \mathrm{Im}\ \xi(q^2),
\end{equation}
and
\begin{equation}
\Lambda_{N0} = - m p_2 \sqrt{1-y_0^2} \left[ M_1-E+\frac{m^2}{4M_1} + \frac{2M_1E-m^2}{2M_1} \mathrm{Re} \, \xi(q^2) + \frac{m^2}{4M_1} \left|\xi(q^2)\right|^2 \right],
\end{equation}
with
\begin{equation}
C_{L0} = -\frac{m^2}{4M_1} [lE_\nu^0+E(l+p_2y_0)].
\end{equation}

An alternative form of the differential decay rate is given by
\begin{equation}
d\Gamma_0 = \frac12 d\Gamma_0^\prime (1 + \hat{\mathbf{s}}_R \cdot \mathbf{P}_0),
\end{equation}
where $\mathbf{P}_0$ is the muon polarization vector implicitly defined in Eq.~(\ref{eq:Pcomp}); explicitly, it reads,
\begin{equation}
\mathbf{P}_0 = P_{L0} \hat{\bm{\epsilon}}_L + P_{T0} \hat{\bm{\epsilon}}_T + P_{N0} \hat{\bm{\epsilon}}_N , \label{eq:p0}
\end{equation}
so the uncorrected components of the muon polarization are simply given by
\begin{equation}
P_{X0} = \frac{\Lambda_{X0}}{a_0^\prime}.
\end{equation}

The three components of $\mathbf{P}_0$ are listed in Table \ref{t:uncP}, where the parameter $\xi(q^2)$ was used in the way described earlier [cf.\ Eq.~(\ref{eq:xiq2}) and the discussion that follows it].

To close this section, it should be remarked that by direct computation, $|\mathbf{P}_0|=1$ at each point in the kinematical region of the Dalitz plot. As a consistency check, similar tables were produced for $\xi(0)=-1,+1,-i,+i$. In all these cases, $|\mathbf{P}_0|=1$ was systematically obtained.

\begingroup
\squeezetable
\begin{table}
\caption{\label{t:uncP}Values of the components of the uncorrected muon polarization vector $\mathbf{P}_0$, Eq.~(\ref{eq:p0}), in the TBR of the process $K_{\mu 3}^+$. The entries correspond to (a) $P_{L0}$, (b) $P_{T0}\times 10^2$, and (c) $P_{N0}$. The energies $E$ and $E_2$ are given in $\textrm{GeV}$. For definiteness, $\mathrm{Re} \, \xi(0) = -0.126$ and $\mathrm{Im} \, \xi(0) = -0.006$ are used.}
\begin{ruledtabular}
\begin{tabular}{lrrrrrrrrrr}
$E_2\backslash E$ & $ 0.1124$ & $ 0.1258$ & $ 0.1392$ & $ 0.1526$ & $ 0.1660$ & $ 0.1794$ & $ 0.1928$ & $ 0.2062$ & $ 0.2196$ & $2330$ \\
\hline
$(a)$    &           &           &           &           &           &           &           &           &           &           \\
$0.2480$ &           &           &           &           &           & $ 0.9968$ & $ 0.9946$ & $ 0.9925$ & $ 0.9888$ & $ 0.9704$ \\
$0.2361$ &           & $ 0.9702$ & $ 0.9578$ & $ 0.9600$ & $ 0.9638$ & $ 0.9666$ & $ 0.9677$ & $ 0.9657$ & $ 0.9558$ & $ 0.8830$ \\
$0.2242$ & $ 0.7592$ & $ 0.8172$ & $ 0.8673$ & $ 0.8984$ & $ 0.9177$ & $ 0.9293$ & $ 0.9345$ & $ 0.9324$ & $ 0.9134$ & $ 0.7347$ \\
$0.2123$ & $ 0.2017$ & $ 0.6004$ & $ 0.7446$ & $ 0.8174$ & $ 0.8585$ & $ 0.8818$ & $ 0.8927$ & $ 0.8901$ & $ 0.8568$ & $ 0.4300$ \\
$0.2004$ & $-0.7359$ & $ 0.2700$ & $ 0.5690$ & $ 0.7064$ & $ 0.7795$ & $ 0.8198$ & $ 0.8383$ & $ 0.8344$ & $ 0.7779$ & $-0.5441$ \\
$0.1885$ &           & $-0.2945$ & $ 0.2972$ & $ 0.5449$ & $ 0.6692$ & $ 0.7351$ & $ 0.7648$ & $ 0.7582$ & $ 0.6605$ &           \\
$0.1766$ &           &           & $-0.1789$ & $ 0.2887$ & $ 0.5046$ & $ 0.6132$ & $ 0.6602$ & $ 0.6477$ & $ 0.4678$ &           \\
$0.1647$ &           &           &           & $-0.1789$ & $ 0.2325$ & $ 0.4224$ & $ 0.4999$ & $ 0.4732$ & $ 0.0944$ &           \\
$0.1528$ &           &           &           &           & $-0.3020$ & $ 0.0821$ & $ 0.2233$ & $ 0.1575$ & $-0.9353$ &           \\
$0.1409$ &           &           &           &           &           & $-0.6953$ & $-0.3670$ & $-0.5862$ &           &           \\
$(b)$    &           &           &           &           &           &           &           &           &           &           \\
$0.2480$ &           &           &           &           &           & $-0.0233$ & $-0.0356$ & $-0.0484$ & $-0.0679$ & $-0.1259$ \\
$0.2361$ &           & $-0.0275$ & $-0.0453$ & $-0.0560$ & $-0.0650$ & $-0.0743$ & $-0.0857$ & $-0.1023$ & $-0.1333$ & $-0.2440$ \\
$0.2242$ & $-0.0398$ & $-0.0655$ & $-0.0783$ & $-0.0878$ & $-0.0968$ & $-0.1070$ & $-0.1207$ & $-0.1421$ & $-0.1843$ & $-0.3519$ \\
$0.2123$ & $-0.0598$ & $-0.0908$ & $-0.1050$ & $-0.1151$ & $-0.1248$ & $-0.1365$ & $-0.1527$ & $-0.1790$ & $-0.2330$ & $-0.4674$ \\
$0.2004$ & $-0.0413$ & $-0.1092$ & $-0.1293$ & $-0.1413$ & $-0.1523$ & $-0.1656$ & $-0.1845$ & $-0.2161$ & $-0.2835$ & $-0.4336$ \\
$0.1885$ &           & $-0.1084$ & $-0.1500$ & $-0.1673$ & $-0.1805$ & $-0.1958$ & $-0.2177$ & $-0.2553$ & $-0.3382$ &           \\
$0.1766$ &           &           & $-0.1544$ & $-0.1908$ & $-0.2096$ & $-0.2279$ & $-0.2535$ & $-0.2979$ & $-0.3975$ &           \\
$0.1647$ &           &           &           & $-0.1960$ & $-0.2359$ & $-0.2612$ & $-0.2920$ & $-0.3440$ & $-0.4470$ &           \\
$0.1528$ &           &           &           &           & $-0.2310$ & $-0.2869$ & $-0.3283$ & $-0.3851$ & $-0.1586$ &           \\
$0.1409$ &           &           &           &           &           & $-0.2067$ & $-0.3130$ & $-0.3155$ &           &           \\
$(c)$    &           &           &           &           &           &           &           &           &           &           \\
$0.2480$ &           &           &           &           &           & $-0.0803$ & $-0.1046$ & $-0.1227$ & $-0.1495$ & $-0.2417$ \\
$0.2361$ &           & $-0.2424$ & $-0.2874$ & $-0.2800$ & $-0.2667$ & $-0.2563$ & $-0.2523$ & $-0.2598$ & $-0.2940$ & $-0.4694$ \\
$0.2242$ & $-0.6509$ & $-0.5764$ & $-0.4978$ & $-0.4392$ & $-0.3972$ & $-0.3695$ & $-0.3559$ & $-0.3614$ & $-0.4072$ & $-0.6784$ \\
$0.2123$ & $-0.9794$ & $-0.7997$ & $-0.6675$ & $-0.5760$ & $-0.5129$ & $-0.4716$ & $-0.4507$ & $-0.4559$ & $-0.5157$ & $-0.9028$ \\
$0.2004$ & $-0.6770$ & $-0.9628$ & $-0.8223$ & $-0.7078$ & $-0.6264$ & $-0.5727$ & $-0.5453$ & $-0.5512$ & $-0.6284$ & $-0.8389$ \\
$0.1885$ &           & $-0.9556$ & $-0.9548$ & $-0.8385$ & $-0.7431$ & $-0.6779$ & $-0.6443$ & $-0.6520$ & $-0.7508$ &           \\
$0.1766$ &           &           & $-0.9838$ & $-0.9574$ & $-0.8634$ & $-0.7899$ & $-0.7510$ & $-0.7619$ & $-0.8838$ &           \\
$0.1647$ &           &           &           & $-0.9838$ & $-0.9726$ & $-0.9064$ & $-0.8661$ & $-0.8809$ & $-0.9955$ &           \\
$0.1528$ &           &           &           &           & $-0.9532$ & $-0.9966$ & $-0.9747$ & $-0.9875$ & $-0.3536$ &           \\
$0.1409$ &           &           &           &           &           & $-0.7186$ & $-0.9302$ & $-0.8101$ &           &           \\
\end{tabular}
\end{ruledtabular}
\end{table}
\endgroup

% --------------------------------
% Virtual RC
% --------------------------------

\section{\label{sec:vrc}Virtual radiative corrections}

The problem of computing RC to the DP of $K_{\mu 3}^\pm$ decays for unpolarized muons to order $\mathcal{O}[(\alpha/\pi)(q/M_1)]$ has been dealt with in full in Ref.~\cite{juarez11}. As it is well-known, there are two types of RC to be accounted for, namely, virtual and bremsstrahlung RC. The virtual RC can be separated into a model-independent part which is finite and calculable and into a model-dependent one which contains the effects of the strong interactions and the intermediate vector boson. Bremsstrahlung RC, on the other hand, can be computed to the same order of approximation by virtue of the Low theorem \cite{low,chew}.

The model-independent transition amplitude with virtual RC to order $\mathcal{O}[(\alpha/\pi)(q/M_1)]$ is provided in Eq.~(33) of Ref.~\cite{juarez11}. It reads,
\begin{equation}
\mathsf{M}_V = \mathsf{M}_0^\prime \left[1 + \frac{\alpha}{2\pi} \Phi_1(E) \right] + \mathsf{M}_{p_1} \frac{\alpha}{2\pi} \Phi_2(E), \label{eq:MV}
\end{equation}
where the functions $\Phi_1(E)$ and $\Phi_2(E)$, and the amplitude $\mathsf{M}_{p_1}$ can be found in Eqs.~(26), (27), and (29) of that reference, respectively.

The quantity $\sum_s |\mathsf{M}_V|^2$ can also be separated into a spin-independent piece and a spin-dependent one. The former has been evaluated in Ref.~\cite{juarez11} and the latter is evaluated here. Now, the case of an emitted polarized muon can be worked out in a close parallelism to the analysis presented in the previous section by introducing again the spin projection operator (\ref{eq:sproj}) in the corresponding spinor of the muon in Eq.~(\ref{eq:MV}). Thus, the differential decay rate of $K_{\mu 3}^+$ decays for polarized emitted muons, including virtual RC, can be cast into the form
\begin{equation}
d\Gamma_V^{(s)} = d\Omega^\prime \mathbf{\hat{s}}_R \cdot \left[ \left(1 + \frac{\alpha}{\pi} \Phi_1 \right) \mathbf{a}_0^{(s)} + \frac{\alpha}{\pi} \Phi_2 \mathbf{a}_V^{(s)} \right], \label{eq:dGV}
\end{equation}
where $\mathbf{a}_0^{(s)}$ has been defined in Eq.~(\ref{eq:a0s}) and
\begin{equation}
\mathbf{a}_V^{(s)} = \frac{E^2}{m^2} \left[ \mathbf{a}_0^{(s)} - \beta a_0^\prime\ \hat{\bm{\epsilon}}_L \right], \label{eq:aV}
\end{equation}
where $\beta=l/E$.

Alternatively, in terms of the orthonormal basis (\ref{eq:basis}), $d\Gamma_V^{(s)}$ reads,
\begin{equation}
d\Gamma_V^{(s)} = d\Omega^\prime \hat{\mathbf{s}}_R \cdot \left( \Lambda_{LV} \hat{\bm{\epsilon}}_L + \Lambda_{TV} \hat{\bm{\epsilon}}_T + \Lambda_{NV} \hat{\bm{\epsilon}}_N \right),
\end{equation}
where
\begin{equation}
\Lambda_{XV} = \Lambda_{X0} \left[1 + \frac{\alpha}{\pi} \left(\Phi_1 + \frac{E^2}{m^2} \Phi_2 \right) \right],
\end{equation}
for $X=N$, $T$, and
\begin{equation}
\Lambda_{LV} = \Lambda_{L0} \left[1 + \frac{\alpha}{\pi} \left(\Phi_1 + \frac{E^2}{m^2} \Phi_2 \right) \right] - \frac{\alpha}{\pi} \frac{E^2}{m^2} \beta a_0^\prime \Phi_2,
\end{equation}

% --------------------------------
% Bremsstrahlung RC 
% --------------------------------

\section{\label{sec:brc}Bremsstrahlung radiative corrections}

The complete analysis of RC to the DP should take into account the emission of a real photon via the process
\begin{equation}
K^+(p_1) \to \pi^0(p_2) + \mu^+ (\ell) + \nu_{\mu} (p_\nu) + \gamma (k), \label{eq:kl3g}
\end{equation}
where $\gamma$ represents a photon with four-momentum $k=(\omega,\mathbf{k})$ and the neutrino four-momentum is now $p_\nu=(E_\nu, \mathbf{p}_\nu)$, so that energy conservation yields $M_1=E_2+E+E_\nu+\omega$.

Following the Low theorem \cite{low,chew}, the decay amplitude for process (\ref{eq:kl3g}) can be organized as
\begin{equation}
\mathsf{M}_B = \sum_{i=1}^3 \mathsf{M}_{B_i}, \label{eq:mb}
\end{equation}
where the different summands in the above equation read
\begin{equation}
\mathsf{M}_{B_1} = -e\mathsf{M}_0 \left[\frac{l\cdot \epsilon}{l\cdot k} - \frac{p_1\cdot \epsilon}{p_1\cdot k}\right], \label{eq:MB1}
\end{equation}
\begin{equation}
\mathsf{M}_{B_2} = -C_K \frac{eG_F}{\sqrt{2}} V_{us}f_+(q^2) \left\{ 2{p_1}_\alpha -[1-\xi(q^2)]q_\alpha \right\} \bar{u}_\nu \mathcal{O}_\alpha \frac{\slashed{k}\slashed{\epsilon}}{2 l\cdot k}v_l, \label{eq:MB2}
\end{equation}
and
\begin{equation}
\mathsf{M}_{B_3} = -C_K \frac{eG_F}{\sqrt{2}} V_{us} f_+(q^2) [1+\xi(q^2)] \left[\frac{p_1 \cdot \epsilon}{p_1\cdot k}k_{\alpha} - \epsilon_\alpha\right] \left[\overline{u}_\nu(p_\nu) O_\alpha v_\ell(l)\right]. \label{eq:MB3}
\end{equation}
The amplitude $\mathsf{M}_{B_1}$ is of order $\mathcal{O}(1/k)$ and contains the infrared divergence, whereas $\mathsf{M}_{B_2}$ and $\mathsf{M}_{B_3}$ are order $\mathcal{O}(k^0)$.

The bremsstrahlung differential decay can be obtained by standard techniques. The explicit expression to start with reads,
\begin{equation}
d\Gamma_B(K^+\to\pi^0\mu^+ \nu_\mu\gamma) = \frac{1}{(2\pi)^8} \frac{1}{2M_1}\frac{mm_\nu}{4E_2EE_\nu\omega} d^3p_2\,d^3l\,d^3p_\nu\,d^3k \, \delta^4(p_1-p_2-l-p_\nu-k) \sum_{\textrm{spins, pol.}}|\mathsf{M}_B|^2, \label{eq:diffdgb}
\end{equation}
where again, the observable effects of spin polarization can then be analyzed through the replacement indicated in relation (\ref{eq:speff}), which in turn allows to separate the spin-independent piece from the spin-dependent one as
\begin{equation}
\sum_{\textrm{spins, pol.}}|\mathsf{M}_B|^2 = \frac12 \sum_{\textrm{spins, pol.}}|\mathsf{M}_B^\prime|^2 - \frac12 \sum_{\textrm{spins, pol.}}|\mathsf{M}_B^{(s)}|^2. \label{eq:sepamplitud}
\end{equation}
The first summand takes part in the analysis of the semileptonic decay of an unpolarized charged kaon dealt with in Ref.~\cite{juarez11}, whereas the individual components $\mathsf{M}_{B_i}^{(s)}$ that make up $\mathsf{M}_B^{(s)}$ can be easily read off from expressions (\ref{eq:MB1})--(\ref{eq:MB3}) once the replacement (\ref{eq:speff}) has been used into them.

A careful analysis in order to properly account for the contribution of the unobserved photons to the Dalitz plot of process (\ref{eq:kl3g}) has been performed in Ref.~\cite{juarez11}. This analysis is quite useful for delimiting the integrations over the kinematical variables in (\ref{eq:diffdgb}). Succinctly, the orientation of the coordinate axes is such that the direction of emission of $\mu^+$ coincides with the $+z$ axis and $\pi^0$ is emitted in the first or fourth quadrant of the $(y,z)$ plane. For this choice, $\hat{\mathbf{p}}_2 \cdot \hat{\mathbf{l}} = \cos\theta_2 \equiv y$, $\hat{\mathbf{l}} \cdot \hat{\mathbf{k}} = \cos\theta_k \equiv x$, and $\hat{\mathbf{p}}_2 \cdot \hat{\mathbf{k}} = \cos\theta_2 \cos\theta_k + \sin\theta_2\sin\theta_k\sin\phi_k$, where $\theta_k$ and $\phi_k$ are the polar and azimuthal angles of the photon and its energy is given by
\begin{equation}
\omega = \frac{F}{2D},
\end{equation}
with
\begin{subequations}
\begin{equation}
F = 2p_2l (y_0-y),
\end{equation}
and
\begin{equation}
D = E_\nu^0 + lx + \mathbf{p}_2 \cdot \hat{\mathbf{k}}.
\end{equation}
\end{subequations}

The TBR of the DP is the region where the three-body decay (\ref{eq:kl3p}) and the four-body decay (\ref{eq:kl3g}) overlap completely. In the TBR the energies $E$ and $E_2$ are restricted to
\begin{equation}
 m\leq E \leq E_m, \qquad E_2^{\mathrm{min}} \leq E_2 \leq E_2^{\mathrm{max}},
\end{equation}
where
\begin{equation}
E_m = \frac{M_1^2-M_2^2+m^2}{2M_1},
\end{equation}
and
\begin{equation}
E_2^{\mathrm{max},\mathrm{min}} = \frac12 (M_1-E\pm l) + \frac{M_2^2}{2(M_1-E \pm l)}.
\end{equation}
The variable $y$, in turn, is restricted to $-1\leq y \leq y_0$.

The FBR is the region where only the four-body decay (\ref{eq:kl3g}) can occur. The energies $E$ and $E_2$ are now restricted to
\begin{equation}
 m\leq E \leq E_c, \qquad M_2 \leq E_2 \leq E_2^{\mathrm{min}},
\end{equation}
where
\begin{equation}
E_c = \frac12 (M_1-M_2) + \frac{m^2}{2(M_1-M_2)}.
\end{equation}
Now, the variable $y$ is restricted to $-1\leq y \leq 1$.

The bremsstrahlung contribution of the FBR is rather simple because the events in that region have the same amplitude $\mathsf{M}_B$ (\ref{eq:mb}) and, most importantly, it is infrared convergent. The differential decay rate in the FBR can be obtained from the one in the TBR by making a few changes in it: The upper limit of the integrals over the variable $y$ becomes one now and the infrared-divergent function $I_0(E,E_2)$ is replaced by the infrared-convergent one
$I_0^F(E,E_2)$ \cite{juarez12}
\begin{equation}
I_0^F = \frac{\theta_0^F}{2} \ln \left( \frac{y_0+1}{y_0-1} \right), \label{eq:i0f}
\end{equation}
where
\begin{equation}
\theta_0^F = 4 \left( \frac{1}{\beta} \mathrm{arctanh} \, \beta -1 \right).
\end{equation}

Following the analysis of the previous section, $d\Gamma_B$ can also be split as
\begin{equation}
d\Gamma_B = \frac12 d\Gamma_B^\prime + \frac12 d\Gamma_B^{(s)}, \label{eq:dGb}
\end{equation}
where $d\Gamma_B^\prime$ and $d\Gamma_B^{(s)}$ are respectively the spin-independent and spin-dependent bremsstrahlung contributions to the DP. The former has been obtained in Ref.~\cite{juarez11} and the latter is the main aim of this section.

\subsection{\label{sec:dgs}Bremsstrahlung RC in the TBR}

% --------------------------------
% dG1
%\subsubsection{$d\Gamma_{B_1}^{(s)}$}
% --------------------------------

After some algebraic manipulations, following the procedure presented in Ref.~\cite{juarez11}, the second term in Eq. \eqref{eq:dGb} can be written as
\begin{equation}
d\Gamma_{B}^{(s)} = \frac{\alpha}{\pi} d\Omega^\prime \hat{\mathbf{s}}_R \cdot \left[ \mathbf{a}_0^{(s)} I_0(E,E_2) + \mathbf{a}_{B}^{(s)} \right]. \label{eq:dG1}
\end{equation}

The first summand in Eq.~(\ref{eq:dG1}) contains the infrared divergence through the term $I_0(E,E_2)$ given in Eq.~(65) of Ref.~\cite{juarez11}. The term $\mathbf{a}_0^{(s)}$ is given in Eq.~(\ref{eq:a0s}). The second summand, $\mathbf{a}_{B}^{(s)}$, comes from the infrared-convergent pieces of $\sum|\mathsf{M}_{B}^{(s)}|^2$, which appears in Eq. \eqref{eq:sepamplitud}, and can be organized as 
\begin{equation}
\mathbf{a}_{B}^{(s)} = \Lambda_{LB} \hat{\bm{\epsilon}}_L + \Lambda_{NB} \hat{\bm{\epsilon}}_N + \Lambda_{TB} \hat{\bm{\epsilon}}_T, \label{eq:aB1c}
\end{equation}
where the different functions $\Lambda_{XB}$ are defined as
\begin{equation}
\Lambda_{XB} = \frac{p_2l}{4\pi} \int_{-1}^{y_0} dy\int_{-1}^1 dx\int_0^{2\pi} d\phi_k \left[ A_X+B_X \mathrm{Re} \, \xi(q^2)+C_X |\xi(q^2)|^2\right], \label{eq:laml1}
\end{equation}
for $X=L,N$, and
\begin{equation}
\Lambda_{TB} = \frac{p_2l}{4\pi} \int_{-1}^{y_0} dy\int_{-1}^1 dx\int_0^{2\pi} d\phi_k D_T \mathrm{Im} \xi(q^2). \label{eq:lambdaT1}
\end{equation}

The explicit forms of $A_L$, $B_L$ and $C_L$, are given by
\begin{eqnarray}
\label{AL}
A_L & = & \frac{1}{(1-\beta x)^2D} \left[ \frac{\beta-x}{8M_1E} \left\{-16M_1^2(\omega^2-E_\nu^0 \omega)+8M_1(E_\nu^0-\omega)[2M_1 E-m^2]-(q^2-m^2)(4M_1^2-m^2) \right\} \right. \nonumber \\
&  & \mbox{} + \left. \frac{\beta(1-x^2)}{4M_1}\left\{4M_1[E(E_\nu^0+E)-M_1(E+\beta l)+p_2ly_0]+q^2(\beta l-E)+D [4M_1(M_1-E-\omega)+m^2]\right\}\right] \nonumber \\
&  & \mbox{} + \frac{1}{(1-\beta x)D}\left[\frac{l (1-x^2)}{4M_1}[q^2+2\omega D+4E_2(M_1+\beta l)-2p_2l(y+y_0)]\right. \nonumber \\
&  & \mbox{} + \frac{\beta-x}{4M_1}\left(4M_1^2\left(2\omega-E_\nu^0\right)-2\omega(E_\nu^0-\omega)\left(8M_1-E\right) +q^2 (M_1+E_2+E_\nu^0) - m^2(M_1+E_2+2\omega) \right. \nonumber \\
&  & \mbox{} + \left. 2D\omega (2M_1-E) + 2l\omega (p_2y+l+\omega x) \right) + \frac{\beta}{4M_1} \left(8M_1^2(E_\nu^0-\omega) -D(4M_1^2+m^2) \right. \nonumber \\
&  & \mbox{} + \left.\left. E(\beta-x)\left\{\left[q^2-m^2+2\omega(D+M_1+E_2)-4M_1E_\nu^0\right]x-2p_2 \omega y\right\}\right)\right] \nonumber \\
&  & \mbox{} + \frac{1}{2M_1D}\left\{l\left[q^2-m^2-6M_1(E_\nu^0 - \omega ) \right] + \omega E(\beta-x)(2E_\nu^0 -3 \omega-D-M_1-E_2)\right\}+l-(1-\beta x) \frac{E\omega l}{M_1D},
\end{eqnarray}
\begin{eqnarray}
\label{BL}
\frac{2M_1El}{m^2} B_L & = & E(E-2M_1) + \frac{4M_1E-E^2-m^2}{1-\beta x} + \frac{m^2-2M_1E}{(1-\beta x)^2} + \frac{E^2(M_1+\beta l+2E+5E_2)}{D} \nonumber \\
&  & \mbox{} - \frac{(1-\beta x)E^2(E+2E_2)}{D} + \frac{(E-2M_1)q^2 - m^2(E+6E_2)+4M_1E(E_\nu^0-E-\beta l)}{2(1-\beta x)D} - 2E\omega \nonumber \\
&  & \mbox{} - \frac{2M_1\omega}{(1-\beta x)^2} + \frac{2E(M_1+2E)\omega}{D} - \frac{2(1-\beta x)E^2\omega}{D} + \frac{(q^2+m^2)(2M_1E-m^2)-4M_1E_\nu^0m^2}{2(1-\beta x)^2DE} \nonumber \\
&  & \mbox{} + \frac{2(M_1+E)\omega}{1-\beta x} - \frac{2(2M_1E+m^2)\omega}{(1-\beta x)D} + \frac{2M_1m^2\omega}{(1-\beta x)^2DE},
\end{eqnarray}
and
\begin{eqnarray}
\label{CL}
C_L & = & \frac{1}{4 (1-\beta x)^2M_1 D E}\left\{ \beta(1-x^2)\left[(D E-q^2)m^2 +(1-\beta x)E^2q^2\right]\right. \nonumber \\
&  & \mbox{} + \left. \frac{1}{2}(\beta-x)\left\{\left(q^2-m^2\right)\left[m^2-2(1-\beta^2 x^2)E^2\right]+2(1-\beta x)E E_\nu^0 m^2\right\}-\beta(1-\beta x)D E m^2\right\}.
\end{eqnarray}

The functions $A_N$, $B_N$, and $C_N$, on the other hand, read
\begin{eqnarray}
\label{AN}
\frac{D}{m} A_N & = & \left\{ \frac{1}{(1-\beta x)^2} \left[ \beta^2 (x^2-1) \frac{4M_1E_2+q^2}{4M_1} + \frac{4M_1^2-m^2}{4M_1} \frac{q^2-m^2}{2E^2} + \frac{\omega}{E^2} \left\{ 2M_1(E-E_\nu^0+\omega) - m^2 \right\} \right. \right. \nonumber \\
&  & \mbox{} \left. + \frac{E_\nu^0}{E^2} (m^2-2M_1E) \right] + \frac{1}{1-\beta x} \left[ \frac{\omega}{2M_1E} \left\{2M_1[4(E_\nu^0-\omega)-2M_1-D] + 2m^2-l^2-lx(M_1-E_2) \right\} \right. \nonumber \\
&  & \mbox{} \left. + \frac{1}{4M_1E} \{(4M_1^2 - q^2)E_\nu^0 - (q^2-m^2)(M_1+E_2+lx)\} \Big] + \frac{\omega}{2M_1} \{ 2(D+E_2) - E_\nu^0 + E + 4\omega\} \right\} \sqrt{1-x^2} \sin \phi_k \nonumber \\
&  & \mbox{} + \frac{\beta^2(1-x^2)}{(1-\beta x)^2M_1} \left[ [M_1 - E(1-\beta x)] p_2 \sqrt{1-y^2} - \frac{D}{4l} [4M_1(M_1-E) + m^2]J \right],
\end{eqnarray}
\begin{eqnarray}
\label{BN}
\frac{D}{m}B_N & = & \left\{\frac{1}{4(1-\beta x)^2M_1}\left[2\beta^2(x^2-1)\left(M_1^2-M_2^2 \right)+(1-\beta^2)\left[q^2-m^2-4M_1(E_\nu^0-\omega)\right] \right] +\frac{\omega}{M_1}\left(D-E_\nu^0-E_2\right) \right. \nonumber \\
&  & \mbox{} + \left.\frac{1}{2(1-\beta x)M_1 E}\left[q^2(E_\nu^0+E_2)-m^2E_2+2M_1(E_\nu^0-\omega)lx-2\omega(m^2+M_1D)\right] \right\}\sqrt{1-x^2} \sin \phi_k \nonumber \\
&  & \mbox{} - \frac{\beta^2(1-x^2)}{2(1-\beta x)^2M_1 l}\left[2\left[M_1-E(1-\beta x)\right]p_2l\sqrt{1-y^2}+\left(2M_1E-m^2\right)DJ\right],
\end{eqnarray}
\begin{eqnarray}
\label{CN}
\frac{4M_1 D}{m} C_N & = & \left[\frac{1}{2(1-\beta x)^2} \left[ 2(x^2-1)\beta^2q^2 + (\beta^2-1)(q^2-m^2) \right] \right. \nonumber \\
&  & \left. + \frac{1}{(1-\beta x)E} \left[ (1+\beta x)(q^2-m^2)E - m^2E_\nu^0\right] \right] \sqrt{1-x^2} \sin \phi_k 
- \frac{(1-\beta^2)(1-x^2)}{(1-\beta x)^2} DJl.
\end{eqnarray}
Finally $D_T$ is given by
\begin{eqnarray}
D_{T} & = & \frac{\beta m}{1-\beta x} \Big\{ \frac{(1-x^2)p_2}{(1-\beta x)D} \Big[ \frac{\sqrt{1-y^2}}{M_1E} \left\{ M_1E - (1-\beta x)E(\omega+E) + \omega [(1-\beta x)E - 2M_1 ] \cos^2 \phi_k \right\} \nonumber \\
&  & \mbox{} - \beta y \sqrt{1-x^2} \sin \phi_k \Big] + \frac{\beta(1-x^2)}{1-\beta x}J - \frac{\sqrt{1-x^2}}{M_1D} [p_2xy(\omega+E-M_1) - E_2 (lx+\omega)] \sin\phi_k \Big\}. \label{eq:lambdaT}
\end{eqnarray}

In the above two expressions the function $J$ was introduced; it reads,
\begin{equation}
J = \frac{\sqrt{1-y^2} - \sqrt{1-y_0^2}}{y_0-y}. \label{eq:J}
\end{equation}

%-------------------------------------------------------------
% Analytical results
%-------------------------------------------------------------

\subsection{\label{sec:ai}Analytical results in the TBR}

Up to this point, all the integrals over the photon variables that make up the different contributions to the bremsstrahlung differential decay rate have been identified and properly defined through the different $\Lambda_{XB}$ functions. These integrals are ready to be performed numerically, so the three components of the muon polarization can be available in a numerical form.

It would be desirable to get the analytical versions of those integrals. However, this is not entirely possible, except for the longitudinal component. For the normal and transverse ones, some elliptic integrals arise and expressing them in terms of elementary functions is beyond the scope of the present paper, so their expressions will be given only in a numerical form.

Thus, the analytical forms of $P_L$ can be given in terms of the functions $\theta_i$ originally introduced in the analysis of RC in baryon semileptonic decays \cite{tun89,tun91,tun93}, and in terms of the additional functions $\eta_0$, $\zeta_{ij}$, $Y_k$, and $Z_k$ defined in Ref.~\cite{torres04}.

The analytical form of $\Lambda_{LB}$ can be organized as
\begin{equation}
\Lambda_{LB} = \lambda_A + \lambda_B \mathrm{Re} \, \xi(q^2) + \lambda_C |\xi(q^2)|^2, \label{eq:lamLA}
\end{equation}
where
\begin{equation}
\lambda_{A,B,C} = \frac{p_2l}{4\pi} \int_{-1}^{y_0} dy \int_{-1}^1 dx \int_0^{2\pi} d\phi_k \left[A_L, B_L, C_L\right]. \label{eq:lambdaABC}
\end{equation}

The explicit integrations of Eq.~(\ref{eq:lambdaABC}) are
\begin{eqnarray}
\frac{2\beta}{p_2l} \lambda_A & = & \left[\left(M_1+\frac{m^2}{4M_1}\right)(2-\beta^2)-2E\right] \theta_0 - \frac{2l^2p_2}{M_1}Y_1 +\left[\frac12(q^2+m^2) - M_1E(1+\beta^2)-\frac{q^2m^2}{4M_1E}\right]Y_2 \nonumber \\
&  & \mbox{} + \left[E\left(3M_1+\frac{q^2}{4M_1}\right)+\left(E+\frac{l^2}{M_1}\right)E_2\right]Y_3-\left(1+\frac{E}{2M_1}\right)Z_2+\frac{m^2}{E}\left[\beta p_2y_0\theta_2+(E+E_\nu^0)(\theta_2-\theta_3)-\beta p_2\theta_{11}\right] \nonumber \\
&  & \mbox{} + M_1 \left[E_\nu^0\left(2-\frac{m^2}{M_1E}\right)+\frac{q^2-m^2}{2E}\left(\frac{m^2}{4M_1^2}-1\right)\right]\left[\theta_3-\frac{m^2}{E^2}\theta_2\right] - \frac{m^2}{E^2} \left[\frac{M_1E_\nu^0}{E}-M_1+\frac{m^2}{2E}\right] \left[2E(\theta_3-\theta_2)+\theta_6\right] \nonumber \\
&  & \mbox{} + \frac{M_1m^2}{E}\left[(3-\beta^2)\theta_2+3\theta_4+\frac{\theta_{24}}{2E^2}\right]+\left[E_\nu^0(2M_1 \beta^2-l\beta)+\frac{l\beta(q^2-m^2)}{4M_1}-2M_1E(3-2\beta^2)-p_2ly_0\right]\theta_3 \nonumber \\
&  & \mbox{} - m^2\left[\left(\frac{1}{E}-\frac{1}{2M_1}\right)\beta p_2y_0+2(3-\beta^2)\right]\theta_3+\left(\theta_4-\frac{m^2}{E^2}\theta_3\right)\left[\frac{q^2}{2}\left(1-\frac{E}{2M_1}\right)-E_\nu^0 M_1-\frac{m^2}{4M_1}(M_1+E_2)\right] \nonumber \\
&  & \mbox{} + \left[E\left(2E_\nu^0-\frac{q^2-m^2}{4M_1}\right)-\frac{m^2\beta p_2y_0}{2M_1}-\frac{q^2+m^2}{2}+l^2\right](\theta_3-\theta_4)+\left[\left(1-\frac{E}{2M_1}\right)p_2ly_0+6m^2+\beta l\left(\frac{q^2-m^2}{4M_1}\right.\right. \nonumber \\
&  & \mbox{} - \left. \left. 2E_\nu^0-\frac{p_2ly_0}{2M_1}\right)\right]\theta_4+\left[\frac{M_1}{E}\left(E_\nu^0-E+\frac{m^2}{2M_1}\right)-\frac{m^2}{E^2}\left(M_1-2E_\nu^0+\frac{q^2-3m^2}{8M_1}\right)-\frac{E}{4M_1}(M_1+E_2)\right. \nonumber \\
&  & \mbox{} + \left. \frac{\beta}{4} \left(l-4M_1\beta+\frac{lE_2}{M_1}+\frac{p_2y_0m^2}{EM_1}\right)\right]\left[2E(\theta_4-\theta_3)+\theta_7\right]-\left(\frac{m^2}{E^2}+\frac{M_1}{2E}\right)\theta_9+\frac{1}{2M_1}\left(2\beta l-E-2M_1\right)\zeta_{10} \nonumber \\
&  & \mbox{} + \left(\frac{m^2}{2M_1E}+2\right)\zeta_{11}+l\left(\frac{p_2ly_0}{M_1}-E_\nu^0+\frac{q^2-m^2}{4M_1}+\frac{6m^2}{E}\right)\theta_5-\frac{l}{2M_1}\left(2E_2+E+l\beta\right)(2l\theta_{10}-\theta_{14}) \nonumber \\
&  & \mbox{} + \left[E(3+4\beta^2)+E_2\left(2-\frac{l\beta}{M_1}\right)-\frac{m^2+E^2}{2M_1}-\frac{\beta^2}{2M_1} (m^2+4M_1^2)-\frac{3l^2}{2M_1}+\frac{(1-y_0)p_2l}{2M_1}\right] \eta_0,
\label{eq:AA}
\end{eqnarray}
\begin{eqnarray}
\frac{4M_1E}{m^2p_2} \lambda_B & = & \left[2(3M_1E+l^2+E^2) - \frac{4M_1E^3}{m^2} \right] \eta_0 + (4M_1E-m^2-E^2)\theta_0 - \frac{2M_1m^2}{E} \left[\frac{q^2+m^2}{4M_1}+M_1-E_2\right] \theta_2 \nonumber \\
&  & \mbox{} + \left[ \frac12(3q^2+m^2)E + m^2(M_1-3E_2) + 2E_\nu^0E(M_1-E) + 2(3-\beta^2)M_1E^2 - 2l^2(M_1+E) \right] \theta_3 \nonumber \\
&  & \mbox{} + E(3EE_2-5M_1E+2l^2-m^2-2p_2ly_0) \theta_4 + El(E+2E_2)\theta_5 - (2M_1E+m^2) \theta_7 \nonumber \\
&  & \mbox{} + \frac{M_1m^2}{E} \theta_6 - 2El^2 \theta_{10} + 2M_1p_2l \theta_{11} + El \theta_{14} + 2E\zeta_{10} - 2(M_1+E) \zeta_{11},
\end{eqnarray}
and
\begin{equation}
\lambda_C = \frac{p_2l}{2\beta} \frac{m^2}{4M_1} \left[ (2-\beta^2) \theta_0 + \frac{m^2(q^2+m^2)}{2E^3} \theta_2 -\frac{(5m^2+q^2)E+2m^2E_\nu^0}{2E^2} \theta_3 + \frac{(E_\nu^0+E)E+m^2}{E} \theta_4 +l \theta_5 - 2\beta^2 \eta_0 \right]. \label{eq:CA}
\end{equation}

Although the above expressions are rather long, they may significantly reduce the computation time required in the numerical evaluations.

\subsection{Bremsstrahlung RC in the FBR}

The FBR of the DP in the semileptonic decay of a kaon plays an important role in the analysis of the polarization of the muon. As it was discussed in the introductory section, the bremsstrahlung RC is a four-body decay whose DP covers entirely the DP of the three-body decay. Even when no experimental arrangement has been made to detect and discriminate real photons, the use of energy conservation makes possible to eliminate the photons that belong to the FBR. Thus, the distinction between these two regions should be clear.

The FBR analysis of the DP proceeds all along in complete analogy with the TBR analysis of the previous section. Therefore, the differential decay rate can also be expressed as
\begin{equation}
d\Gamma_B^F = \frac12 d\Gamma_B^{\prime F} + \frac12 d\Gamma_B^{(s) F}. \label{eq:DGFB}
\end{equation}
Hereafter, the superscript $F$ attached to a quantity $W$, \textit{i.e.}, $W^F$, will be used as a reminder that $W$ is defined in the FBR.

The first summand in Eq.~(\ref{eq:DGFB}) corresponds to the spin-independent part, which has already been reported in Ref.~\cite{juarez12}, while the second one is spin dependent and can be obtained from the $d\Gamma_{B}^{(s)}$, Eq.~(\ref{eq:dG1}), by performing the simple changes indicated in the previous sections. Therefore,
\begin{equation}
d\Gamma_{B}^{(s) F} = \frac{\alpha}{\pi} d\Omega^\prime \hat{\mathbf{s}}_R \cdot \left[ \mathbf{a}_0^{(s) F} I_0^F(E,E_2) + \mathbf{a}_{B}^{(s) F} \right], \label{eq:dG1F}
\end{equation}
where
\begin{equation}
\mathbf{a}_0^{(s) F} = \Lambda_{L0} \hat{\bm{\epsilon}}_L, \label{eq:a0F}
\end{equation}
the function $I_0^F$ in given in Eq.~(\ref{eq:i0f}), and the quantity $\mathbf{a}_{B}^{(s) F}$ has a similar form to $\mathbf{a}_{B}^{(s)}$, i.e.,
\begin{equation}
\mathbf{a}_{B}^{(s) F} = \Lambda_{L}^F \hat{\bm{\epsilon}}_L + \Lambda_{N}^F \hat{\bm{\epsilon}}_N + \Lambda_{T}^F \hat{\bm{\epsilon}}_T. \label{eq:aB1F}
\end{equation}

It should be remarked that the explicit forms of $\Lambda_{X}^F$, $X=L,T,N$, can be obtained from their counterparts $\Lambda_{XB}$ defined in the TBR listed in the previous section by simply replacing the upper limit of integration over the $y$ variable with one and performing the change
\begin{equation}
J\rightarrow J^F=-\frac{\sqrt{1-y^2}}{y_0-y}, \label{eq:JJF}
\end{equation}
directly during the integration process of $A_N$, $B_N$, $C_N$ and $D_T$.

\subsubsection{Analytical results for the FBR}

Likewise the TBR case, only analytical results for the longitudinal component are available because again the normal and transverse ones are not possible to be integrated by using ordinary methods. The analytical version of $\Lambda_L^F$ defined in Eq.~(\ref{eq:aB1F}) can be obtained from its counterpart of the TBR through the changes indicated in the previous sections. Thus, in a total analogy with Eq.~(\ref{eq:lamLA}), $\Lambda_L^F$ can be expressed as
\begin{equation}
\Lambda_L^F = \lambda_A^F + \lambda_B^F \mathrm{Re} \, \xi(q^2) + \lambda_C^F |\xi(q^2)|^2, \label{eq:lamLAF}
\end{equation}
with
\begin{equation}
\lambda_{A,B,C}^F = \frac{p_2l}{4\pi}\int_{-1}^{1} dy \int_{-1}^1 dx \int_0^{2\pi} d\phi_k [A_L, B_L, C_L],
\end{equation}
and their explicit forms can be obtained from the corresponding $\lambda_{A,B,C}$ given in Eqs.~(\ref{eq:AA})--(\ref{eq:CA}) simply by attaching a superscript $F$ to the $\theta_i$, $\zeta_{ij}$, $Y_k$, $Z_k$, and $\eta_0$ functions. Additionally, for $\lambda_{A}^F$ it is necessary to eliminate the contribution $(1-y_0)p_2l$ that comes along $\eta_0$ in the very last summand of Eq.~(\ref{eq:AA}).

\section{\label{sec:DPComplete}Complete Dalitz plot}

The differential decay rate of $K_{\mu 3}^+$ decays in the variables $E$ and $E_2$, that is, the DP, with nonzero polarization of the emitted muon and including radiative corrections to order $\mathcal{O}[(\alpha/\pi)(q/M_1)]$ is given by
\begin{eqnarray}
d\Gamma(K_{\mu 3}^+) & = & d\Gamma_V + d\Gamma_B \nonumber \\
& = & \frac12 \left[ d\Gamma_V^\prime + d\Gamma_B^\prime + {d\Gamma_B^\prime}^F \right] + \frac12 \left[ d\Gamma_V^{(s)} + d\Gamma_B^{(s)} + d\Gamma_B^{(s) F} \right], \label{eq:fulldp}
\end{eqnarray}
where the primed quantities summarize the unpolarized case already discussed \cite{juarez11,juarez12} and the remaining quantities, labeled with the superscript $(s)$, originate from the muon polarization itself. In this regard, $d\Gamma_V^{(s)}$ is given in Eq.~(\ref{eq:dGV}) and the bremsstrahlung counterpart is constituted by TBR and FBR pieces, Eqs.~(\ref{eq:dG1}) and (\ref{eq:dG1F}), respectively.

The triple integration over the real photon variables was performed analytically in some limited cases, but in the others it remains to be performed numerically. It is not a flaw of the analysis; ultimately, modern numerical methods can provide accurate solutions to all the standing integrals. On the other hand, the infrared divergence and the finite terms that accompany it have been explicitly and analytically extracted, however.

Despite its length, the form of Eq.~(\ref{eq:fulldp}) is basically simple and organized in a way that is easy to handle. Its main usefulness lies in that it can be used to evaluate the effects of RC on the muon polarization, as it is described in the next section.

\section{\label{sec:mpol}The muon polarization vector with RC}

At this stage, all the partial results can be gathered together to construct the muon polarization vector with RC, hereafter denoted by $\mathbf{P}$, which is a function of $E$ and $E_2$. Therefore,
\begin{equation}
\mathbf{P} = \mathbf{P}_0 + \mathbf{P}_{RC}, \label{eq:ptot}
\end{equation}
where $\mathbf{P}_0$ is the uncorrected muon polarization vector introduced in Eq.~(\ref{eq:p0}) and $\mathbf{P}_{RC}$ collects both virtual and bremsstrahlung RC contributions to $\mathbf{P}_0$. It can be defined as \cite{gins71}
\begin{equation}
\mathbf{P}_{RC} = \frac{\alpha}{\pi} \frac{\mathbf{a}^{(s)}-a\mathbf{P_0}}{a_0^\prime + (\alpha/\pi) a}, \label{eq:prc}
\end{equation}
where
\begin{equation}
\mathbf{a}^{(s)} = (\Phi_1+\theta_1) \mathbf{a}_0^{(s)} + \Phi_2 \mathbf{a}_V^{(s)} + \mathbf{a}_B^{(s)}, \label{eq:as}
\end{equation}
and
\begin{equation}
a = (\Phi_1+\theta_1) a_0^\prime + \Phi_2 a_V^\prime + a_B^\prime. \label{eq:a}
\end{equation}

The vectors that constitute $\mathbf{a}^{(s)}$, namely, $\mathbf{a}_0^{(s)}$, $\mathbf{a}_V^{(s)}$, and $\mathbf{a}_B^{(s)}$, originate from spin-dependent contributions to the decay amplitude. The first one is of course the uncorrected term defined in Eq.~(\ref{eq:a0s}) whereas the last two are the virtual and bremsstrahlung RC contributions defined respectively in Eqs.~(\ref{eq:aV}) and (\ref{eq:aB1c}). The term $\Phi_1+\theta_1$ and $\Phi_2$ are provided in Ref.~\cite{juarez11}. $\Phi_1$ and $\theta_1$ encode separately infrared divergent terms but $\Phi_1+\theta_1$ is finite.

On the other hand, the spin-independent term $a$ in Eq.~(\ref{eq:a}) is written in terms of $a_0^\prime$ which is defined in Eq.~(\ref{eq:Amp0sin}), and
\begin{equation}
a_V^\prime = \frac{M_1^3}{8} \left[ A_1^{(V)} + A_2^{(V)} \mathrm{Re} \, \xi(q^2) + A_3^{(V)} |\xi(q^2)|^2 \right],
\end{equation}
and
\begin{equation}
a_B^\prime = \frac{M_1^3}{8} \left[ A_1^{(B)} + A_2^{(B)} \mathrm{Re} \, \xi(q^2) + A_3^{(B)} |\xi(q^2)|^2 \right],
\end{equation}
where $A_m^{(V)}$ and $A_m^{(B)}$ $m=1,2,3$, are defined in Ref.~\cite{juarez11}.

The magnitude of the muon polarization vector is obtained in the usual way as
\begin{equation}
P = |\mathbf{P}| = \sqrt{(P_{L0}+P_{LRC})^2+(P_{T0}+P_{TRC})^2+(P_{N0}+P_{NRC})^2}, \label{eq:pola}
\end{equation}
where $P_{X0}$ and $P_{XRC}$, $X=L,T,N$, are the longitudinal, transverse, and normal components of $\mathbf{P_0}$ and $\mathbf{P_{RC}}$, respectively.

With all the inputs properly defined, the numerical evaluation of the RC to the components of the muon polarization, Eq.~(\ref{eq:prc}), and the magnitude $P$, can be performed at any point in the allowed kinematical region. Samples of these numbers are displayed in Tables \ref{t:crP} and \ref{t:crPMag}, respectively. Particularly, notice in Table \ref{t:crPMag} that at each point of the kinematical region, $P \leq 1$, as required by unitarity. As a consistency check, $P_0$ and $P$ were also evaluated for $\xi(q^2)=1$, $-1$, $i$, and $-i$. In all these scenarios, $P_0=1$ and $P\leq 1$ were observed.

\begingroup
\squeezetable
\begin{table}
\caption{\label{t:crP}RC to the components of the muon polarization vector $\mathbf{P}_0$, Eq.~(\ref{eq:prc}), in the TBR of the process $K_{\mu 3}^+$. The entries correspond to (a) $P_{LRC} \times 10^2$, (b) $P_{TRC} \times 10^4$ and (c) $P_{NRC} \times 10^2$. The energies $E$ and $E_2$ are given in $\textrm{GeV}$. For definiteness, $\mathrm{Re} \, \xi(0) = -0.126$ and $\mathrm{Im} \, \xi(0) = -0.006$ are used.}
\begin{ruledtabular}
\begin{tabular}{lrrrrrrrrrr}
$E_2\backslash E$    & $0.1124$ & $0.1258$ & $0.1392$ & $0.1526$ & $0.1660$ & $0.1794$ & $0.1928$ & $0.2062$ & $0.2196$ & $2330$ \\
\hline
$(a)$    &           &           &           &           &           &           &           &           &           &           \\
$0.2480$ &           &           &           &           &           & $ 0.0001$ & $ 0.0019$ & $ 0.0037$ & $ 0.0056$ & $ 0.0105$ \\
$0.2361$ &           & $-0.0004$ & $ 0.0060$ & $ 0.0104$ & $ 0.0128$ & $ 0.0139$ & $ 0.0146$ & $ 0.0153$ & $ 0.0172$ & $ 0.0303$ \\
$0.2242$ & $ 0.0181$ & $ 0.0368$ & $ 0.0386$ & $ 0.0365$ & $ 0.0335$ & $ 0.0307$ & $ 0.0286$ & $ 0.0277$ & $ 0.0298$ & $ 0.0567$ \\
$0.2123$ & $ 0.0939$ & $ 0.0978$ & $ 0.0834$ & $ 0.0696$ & $ 0.0587$ & $ 0.0506$ & $ 0.0449$ & $ 0.0420$ & $ 0.0446$ & $ 0.0925$ \\
$0.2004$ & $ 0.1692$ & $ 0.1793$ & $ 0.1425$ & $ 0.1122$ & $ 0.0902$ & $ 0.0748$ & $ 0.0645$ & $ 0.0590$ & $ 0.0624$ & $ 0.0737$ \\
$0.1885$ &           & $ 0.2786$ & $ 0.2221$ & $ 0.1686$ & $ 0.1309$ & $ 0.1052$ & $ 0.0885$ & $ 0.0796$ & $ 0.0838$ &           \\
$0.1766$ &           &           & $ 0.3268$ & $ 0.2453$ & $ 0.1847$ & $ 0.1443$ & $ 0.1183$ & $ 0.1044$ & $ 0.1080$ &           \\
$0.1647$ &           &           &           & $ 0.3459$ & $ 0.2563$ & $ 0.1944$ & $ 0.1548$ & $ 0.1325$ & $ 0.1238$ &           \\
$0.1528$ &           &           &           &           & $ 0.3353$ & $ 0.2515$ & $ 0.1928$ & $ 0.1535$ & $ 0.0132$ &           \\
$0.1409$ &           &           &           &           &           & $ 0.2285$ & $ 0.1749$ & $ 0.0844$ &           &           \\
$(b)$    &           &           &           &           &           &          &            &           &           &           \\
$0.2480$ &           &           &           &           &           & $ 0.0069$ & $ 0.0110$ & $ 0.0147$ & $ 0.0187$ & $ 0.0256$ \\
$0.2361$ &           & $ 0.0038$ & $ 0.0084$ & $ 0.0122$ & $ 0.0154$ & $ 0.0180$ & $ 0.0204$ & $ 0.0229$ & $ 0.0261$ & $ 0.0351$ \\
$0.2242$ & $ 0.0025$ & $ 0.0087$ & $ 0.0136$ & $ 0.0173$ & $ 0.0201$ & $ 0.0224$ & $ 0.0244$ & $ 0.0266$ & $ 0.0301$ & $ 0.0401$ \\
$0.2123$ & $ 0.0023$ & $ 0.0104$ & $ 0.0162$ & $ 0.0202$ & $ 0.0230$ & $ 0.0251$ & $ 0.0270$ & $ 0.0292$ & $ 0.0329$ & $ 0.0352$ \\
$0.2004$ & $-0.0024$ & $ 0.0086$ & $ 0.0164$ & $ 0.0213$ & $ 0.0244$ & $ 0.0266$ & $ 0.0285$ & $ 0.0307$ & $ 0.0344$ & $-0.0110$ \\
$0.1885$ &           & $-0.0006$ & $ 0.0123$ & $ 0.0196$ & $ 0.0239$ & $ 0.0266$ & $ 0.0287$ & $ 0.0309$ & $ 0.0335$ &           \\
$0.1766$ &           &           & $-0.0020$ & $ 0.0124$ & $ 0.0198$ & $ 0.0240$ & $ 0.0267$ & $ 0.0288$ & $ 0.0281$ &           \\
$0.1647$ &           &           &           & $-0.0096$ & $ 0.0077$ & $ 0.0160$ & $ 0.0205$ & $ 0.0223$ & $ 0.0117$ &           \\
$0.1528$ &           &           &           &           & $-0.0275$ & $-0.0048$ & $ 0.0049$ & $ 0.0064$ & $-0.0108$ &           \\
$0.1409$ &           &           &           &           &           & $-0.0613$ & $-0.0290$ & $-0.0192$ &           &           \\
$(c)$    &           &           &           &           &           &           &           &           &           &           \\
$0.2480$ &           &           &           &           &           & $ 0.0286$ & $ 0.0385$ & $ 0.0429$ & $ 0.0454$ & $ 0.0515$ \\
$0.2361$ &           & $ 0.0438$ & $ 0.0730$ & $ 0.0812$ & $ 0.0813$ & $ 0.0780$ & $ 0.0734$ & $ 0.0688$ & $ 0.0656$ & $ 0.0719$ \\
$0.2242$ & $ 0.0703$ & $ 0.1139$ & $ 0.1209$ & $ 0.1166$ & $ 0.1084$ & $ 0.0993$ & $ 0.0904$ & $ 0.0826$ & $ 0.0775$ & $ 0.0820$ \\
$0.2123$ & $ 0.0965$ & $ 0.1432$ & $ 0.1474$ & $ 0.1391$ & $ 0.1270$ & $ 0.1144$ & $ 0.1028$ & $ 0.0928$ & $ 0.0859$ & $ 0.0717$ \\
$0.2004$ & $ 0.0261$ & $ 0.1381$ & $ 0.1580$ & $ 0.1526$ & $ 0.1398$ & $ 0.1255$ & $ 0.1119$ & $ 0.1000$ & $ 0.0907$ & $-0.0208$ \\
$0.1885$ &           & $ 0.0657$ & $ 0.1425$ & $ 0.1536$ & $ 0.1454$ & $ 0.1316$ & $ 0.1171$ & $ 0.1034$ & $ 0.0896$ &           \\
$0.1766$ &           &           & $ 0.0621$ & $ 0.1281$ & $ 0.1372$ & $ 0.1290$ & $ 0.1155$ & $ 0.1001$ & $ 0.0765$ &           \\
$0.1647$ &           &           &           & $ 0.0292$ & $ 0.0966$ & $ 0.1078$ & $ 0.1005$ & $ 0.0838$ & $ 0.0358$ &           \\
$0.1528$ &           &           &           &           & $-0.0403$ & $ 0.0403$ & $ 0.0552$ & $ 0.0397$ & $-0.0241$ &           \\
$0.1409$ &           &           &           &           &           & $-0.1653$ & $-0.0564$ & $-0.0423$ &           &           \\
\end{tabular}
\end{ruledtabular}
\end{table}
\endgroup

\begingroup
\squeezetable
\begin{table}
\caption{\label{t:crPMag}Magnitude of the muon polarization vector $P$, Eq.~(\ref{eq:pola}), in the TBR of the process $K_{\mu 3}^+$. The energies $E$ and $E_2$ are given in $\textrm{GeV}$. For definiteness, $\mathrm{Re} \, \xi(0) = -0.126$ and $\mathrm{Im} \, \xi(0) = -0.006$ are used.}
\begin{ruledtabular}
\begin{tabular}{lrrrrrrrrrr}
$E_2\backslash E$ & $ 0.1124$ & $ 0.1258$ & $ 0.1392$ & $ 0.1526$ & $ 0.1660$ & $ 0.1794$ & $ 0.1928$ & $ 0.2062$ & $ 0.2196$ & $2330$ \\
\hline
$0.2480$ &           &           &           &           &           & $ 1.0000$ & $ 1.0000$ & $ 1.0000$ & $ 1.0000$ & $ 1.0000$ \\
$0.2361$ &           & $ 0.9999$ & $ 0.9999$ & $ 0.9999$ & $ 1.0000$ & $ 1.0000$ & $ 1.0000$ & $ 1.0000$ & $ 1.0000$ & $ 1.0000$ \\
$0.2242$ & $ 0.9997$ & $ 0.9997$ & $ 0.9998$ & $ 0.9999$ & $ 0.9999$ & $ 1.0000$ & $ 1.0000$ & $ 1.0000$ & $ 1.0000$ & $ 0.9999$ \\
$0.2123$ & $ 0.9993$ & $ 0.9995$ & $ 0.9997$ & $ 0.9998$ & $ 0.9999$ & $ 1.0000$ & $ 1.0000$ & $ 1.0000$ & $ 1.0000$ & $ 0.9998$ \\
$0.2004$ & $ 0.9986$ & $ 0.9992$ & $ 0.9996$ & $ 0.9998$ & $ 0.9999$ & $ 0.9999$ & $ 1.0000$ & $ 1.0000$ & $ 1.0000$ & $ 0.9998$ \\
$0.1885$ &           & $ 0.9986$ & $ 0.9994$ & $ 0.9997$ & $ 0.9998$ & $ 0.9999$ & $ 1.0000$ & $ 1.0000$ & $ 0.9999$ &           \\
$0.1766$ &           &           & $ 0.9989$ & $ 0.9995$ & $ 0.9998$ & $ 0.9999$ & $ 1.0000$ & $ 1.0000$ & $ 0.9999$ &           \\
$0.1647$ &           &           &           & $ 0.9991$ & $ 0.9997$ & $ 0.9999$ & $ 1.0000$ & $ 0.9999$ & $ 0.9998$ &           \\
$0.1528$ &           &           &           &           & $ 0.9994$ & $ 0.9999$ & $ 0.9999$ & $ 0.9999$ & $ 1.0000$ &           \\
$0.1409$ &           &           &           &           &           & $ 0.9997$ & $ 0.9999$ & $ 0.9999$ &           &           \\
\end{tabular}
\end{ruledtabular}
\end{table}
\endgroup

\subsection{RC to the polarization vector for the muon energy spectrum}

\begin{figure}[ht]
\scalebox{0.98}{\includegraphics{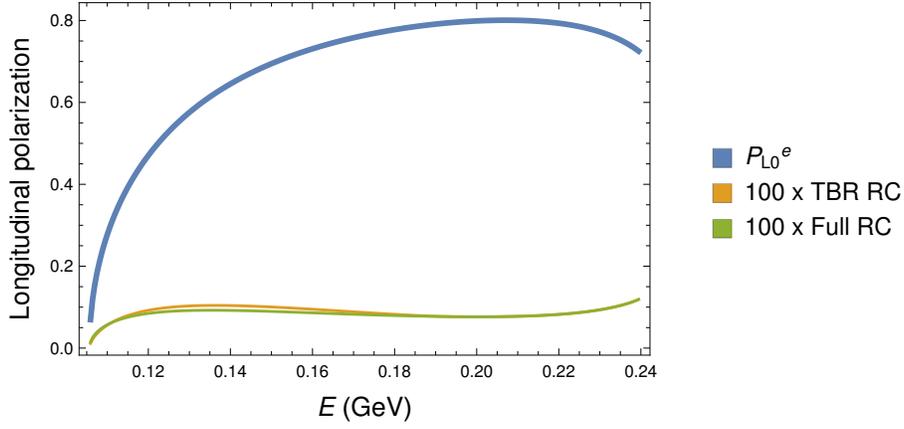}}
\caption{\label{fig:pl}Longitudinal component of the polarization vector for the muon energy spectrum, $P_{L0}^e$, along with the RC from the TBR and TBR+FBR; the latter are labeled as \lq\lq Full RC". Both types of RC are multiplied by 100. For definiteness, $\mathrm{Re} \, \xi(0) = -0.126$ and $\mathrm{Im} \, \xi(0) = -0.006$ are used.}
\end{figure}

\begin{figure}[ht]
\scalebox{0.98}{\includegraphics{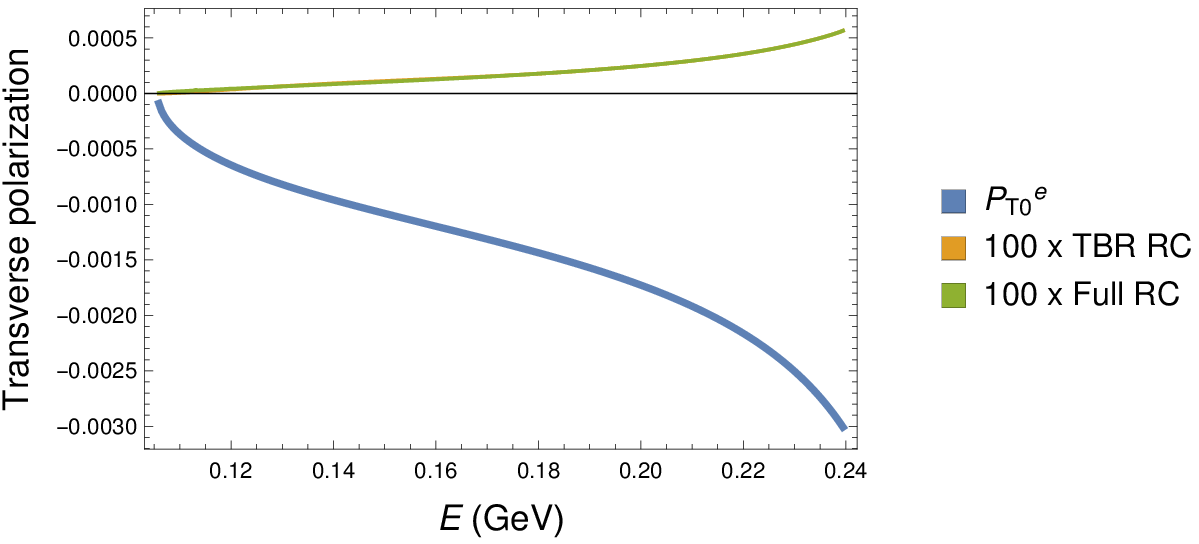}}
\caption{\label{fig:pt}Transverse component of the polarization vector for the muon energy spectrum, $P_{T0}^e$, along with the RC from the TBR and TBR+FBR; the latter are labeled as \lq\lq Full RC". Both types of RC are multiplied by 100. For definiteness, $\mathrm{Re} \, \xi(0) = -0.126$ and $\mathrm{Im} \, \xi(0) = -0.006$ are used.}
\end{figure}

\begin{figure}[ht]
\scalebox{0.98}{\includegraphics{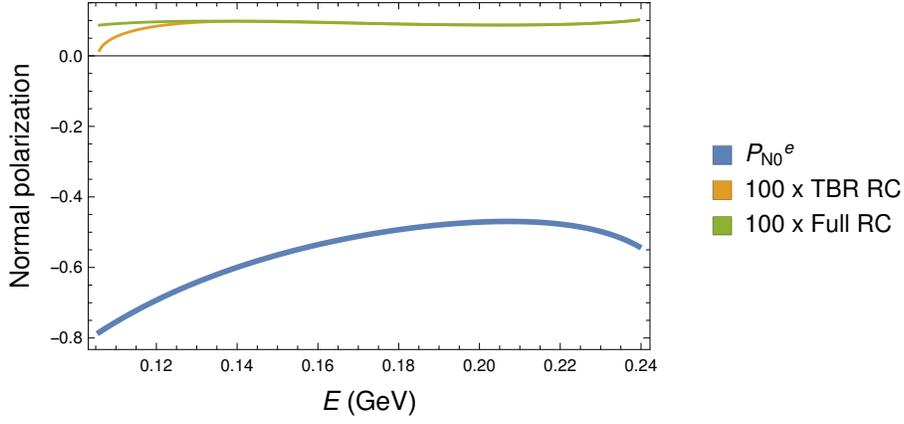}}
\caption{\label{fig:pn}Normal component of the polarization vector for the muon energy spectrum, $P_{N0}^e$, along with the RC from the TBR and TBR+FBR; the latter are labeled as \lq\lq Full RC". Both types of RC are multiplied by 100. For definiteness, $\mathrm{Re} \, \xi(0) = -0.126$ and $\mathrm{Im} \, \xi(0) = -0.006$ are used.}
\end{figure}

At the DP level, the RC from the FBR do not participate into the muon polarization. In order to appreciate the strict participation of this FBR, a further integration step over the pion energy $E_2$ can be performed to end up with a polarization vector for the muon energy spectrum, denoted here by $\mathbf{P}^e(E)$. Thus
\begin{equation}
\mathbf{P}^e = \mathbf{P}_0^e + \mathbf{P}_{RC}^e, \label{eq:pe}
\end{equation}
where the components of the uncorrected polarization vector are straightforwardly defined as
\begin{equation}
P_{X0}^e = \frac{\displaystyle \int_{E_2^\mathrm{min}}^{E_2^\mathrm{max}} dE_2 \Lambda_{X0}}{\displaystyle \int_{E_2^\mathrm{min}}^{E_2^\mathrm{max}} dE_2 a_0^\prime}, \label{eq:p0e}
\end{equation}
and the radiatively corrected version is given by
\begin{equation}
\mathbf{P}_{RC}^e = \frac{\alpha}{\pi} \frac{\displaystyle \int_{E_2^\mathrm{min}}^{E_2^\mathrm{max}} dE_2 \mathbf{a}^{(s)} + \int_{M_2}^{E_2^\mathrm{min}} dE_2 \left[ \mathbf{a}_0^{(s)F}I_0^F+\mathbf{a}_B^{(s)F} \right] - \left[ \int_{E_2^\mathrm{min}}^{E_2^\mathrm{max}}dE_2 a + \int_{M_2}^{E_2^{min}} dE_2 a_B^F \right] \mathbf{P}_0^e}{ \displaystyle \int_{E_2^\mathrm{min}}^{E_2^\mathrm{max}} dE_2 \left[a_0^\prime + \frac{\alpha}{\pi}a\right] + \frac{\alpha}{\pi}\int_{M_2}^{E_2^\mathrm{min}} dE_2 a_B^F}, \label{eq:prce}
\end{equation}
where $\mathbf{a}_B^{(s) F}$ is introduced in Eq.~(\ref{eq:a0F}) and
\begin{equation}
a_B^F = \frac{M_1^3}{8} \left[ A_{1F} + A_{2F} \mathrm{Re} \, \xi(q^2) + A_{3F} |\xi(q^2)|^2 \right],
\end{equation}
with $A_{iF}$ given in Eq.~(12) of Ref.~\cite{juarez12}. The longitudinal, transverse, and normal components of $\mathbf{P}_0^e$ along with their respective RC are plotted in Figs.~\ref{fig:pl}--\ref{fig:pn} for the sake of completeness.

A final step of integration over the muon energy $E$ can be performed on the numerators and denominators of Eqs.~(\ref{eq:p0e}) and (\ref{eq:prce}) to obtain the totally integrated components of the muon polarization, $P_X^\mathrm{tot}$, in a close analogy to Eq.~(\ref{eq:pe}). With all the necessary inputs, the numerical values are found to be
\begin{subequations}
\label{eq:ptotfi}
\begin{eqnarray}
P_L^\mathrm{tot} = 0.7040 + \left\{ \begin{array}{rl} 0.0007 & \mbox{(TBR)} \\ 0.0006 & \mbox{(TBR+FBR)}, \end{array} \right.
\end{eqnarray}
\begin{eqnarray}
P_T^\mathrm{tot} = -1.296 \times 10^{-3} + \left\{ \begin{array}{rl} 0.003 \times 10^{-3} & \mbox{(TBR)} \\ 0.003 \times 10^{-3} & \mbox{(TBR+FBR)}, \end{array} \right.
\end{eqnarray}
and
\begin{eqnarray}
P_N^\mathrm{tot} = -0.5445 + \left\{ \begin{array}{rl} 0.0008 & \mbox{(TBR)} \\ 0.0008 & \mbox{(TBR+FBR)}, \end{array} \right.
\end{eqnarray}
\end{subequations}
where the first summand in each one of Eqs.~(\ref{eq:ptotfi}) is the uncorrected value and the second summands are either RC from the TBR (first line) or the full FBR+TBR (second line). In the latter case, notice that the effects of the FBR alone cannot be disentangled. In particular, the $P_T^\mathrm{tot}$ value is of the order of the measured one suggested by the Particle Data Group, $P_T = (-1.7 \pm 2.3 \pm 1.1)\times 10^{-3}$ \cite{part}.

\begin{figure}[ht]
\scalebox{0.98}{\includegraphics{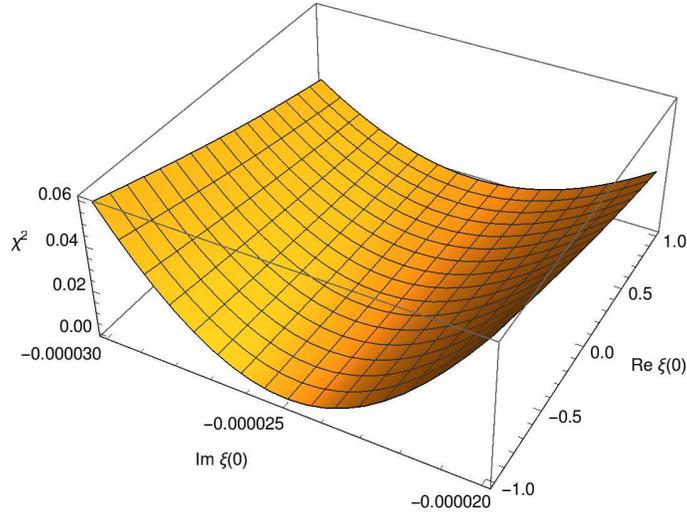}}
\caption{\label{fig:x2plot}$\chi^2$ distribution plot using $P_T$ as a function of the real and imaginary parts of $\xi(0)$.
}
\end{figure}

\section{Application: A $\chi^2$ distribution with two degrees of freedom}

A $\chi^2$ distribution with two degrees of freedom, to wit, the real and imaginary parts of $\xi(0)$, can be evaluated using the formulas obtained for the transverse polarization with radiative corrections as obtained in the previous section. For this purpose, the function tested was constructed as
\begin{equation}
\chi^2(\mathrm{Re} \, \xi_k,\mathrm{Im} \, \xi_{k^\prime}) = \frac{1}{n} \sum_{i=1}^{n}\ \left|P_{Ti}(M_{1i},M_{2i},m_{i},\mathrm{Re} \, \xi_{k},\mathrm{Im} \, \xi_{k^\prime}) - P_{Ti}^\mathrm{exp} \right|^2,
\end{equation}
where the observed transverse polarization $P_T$ and masses $M_1$, $M_2$, and $m$ were randomly generated around their central values and within their uncertainties, assuming in each case a normal distribution.

The procedure followed is simple. First, an $n\times n$ lattice was generated, so the function $\chi^2$ was evaluated at the points $(\mathrm{Re} \, \xi_k, \mathrm{Im} \, \xi_{k^\prime})$, within the intervals $-1 \leq \mathrm{Re} \, \xi_k \leq 1$ and $-1 \leq \mathrm{Im} \, \xi_{k^\prime} \leq 1$, for $k,k^\prime=1,\ldots,n$. As a case example, $n=20$ was set. 

The distribution plot of $\chi^2$ is displayed in Fig.~\ref{fig:x2plot}. A close inspection reveals that $\chi^2$ is a convex function with no global minimum. Instead, the minima are inside a narrow, parabolic shaped flat valley at roughly $\mathrm{Im} \, \xi(0) \approx -0.000025$, which, for practical purposes, is consistent with zero.

The overall analysis is consistent with expectations in the sense that the theoretical prediction is comparable to the observed one.

\section{\label{sec:closing}Discussion and concluding remarks}

In the present paper, the differential decay rate of the process $K_{\mu 3}^+$, in the variables $E$ and $E_2$, has been obtained. This so-called DP comprises RC of order $\mathcal{O}[(\alpha/\pi)(q/M_1)]$, where $q$ is the momentum transfer and $M_1$ is the mass of the decaying kaon. RC are accounted for in both the TBR and FBR of the kinematically allowed region. The FBR must be incorporated into the RC when bremsstrahlung photons cannot be discriminated either kinematically or by direct detection. A novelty in the analysis is that the muon is considered to be polarized, which complements the unpolarized case discussed in previous works \cite{juarez11,juarez12}.

From $d\Gamma(K_{\mu 3}^+)$, Eq.~(\ref{eq:fulldp}), the muon polarization vector $\mathbf{P}$ in the variables $E$ and $E_2$ with RC can be straightforwardly obtained. It is given by Eq.~(\ref{eq:ptot}).

The RC to the uncorrected $P_{0}$, corresponding to the TBR, are listed in Table \ref{t:crP} for each one of its components, namely $P_{L}$, $P_{T}$, and $P_{N}$. Relative to the uncorrected value, the RC to $P_{LRC}$ are around $0.3\%$ near the bottom of the DP, and decrease bottom upwards up to $0.005\%$. For $P_{NRC}$, RC increase from $0.05\%$ at the upper border, until they reach a maximun value of around $0.15\%$ roughly at the middle of the $E_{2}$ values, then they decrease up to reaching negative values, around $-0.06\%$. For $P_{TRC}$ a similar behavior is observed, but there, the values are very small, at the level of $10^{-5}$ down to $10^{-6}$. It can thus be concluded that while the RC for $P_L$ and $P_N$ are of order $\sim 10^{-2}$, the RC for $P_T$ can practically be neglected. This conclusion can not be reached \textit{a priori}. After a long but otherwise standard calculation this statement was possible. Of course, these conclusions are reached for the values $\mathrm{Re} \, \xi(0) = -0.126$ and $\mathrm{Im} \, \xi(0) = -0.006$, which are the closest to physics \cite{part}. Different values might yield different scenarios.

On the other hand, the magnitude of the polarization vector $P$, Eq.~(\ref{eq:pola}), evaluated at various points in the DP and listed in Table \ref{t:crPMag}, yields some interesting results. First, notice that the RC are negligibly small at the upper-right corner of the DP and gradually increase their effects on $P$ so it {\it decreases} from top to bottom and from right to left. The highest correction obtained on $P$ is around $0.15\%$ at the lower-left corner of the DP. Secondly, on general grounds, RC affects more the direction rather than the magnitude of the polarization vector of the muon.

As for the effects of the FBR on $\mathbf{P}$, they can be better seen in Figs.~\ref{fig:pl}-\ref{fig:pn} where the RC to the polarization vector for the muon energy spectrum are plotted. In this case, the RC are always positive and also amount a fraction of a percent. While in $P_L$ the FBR effects are noticeable, in $P_T$ and $P_N$ they are barely significant, so the TBR is more relevant.

Finally, the fully integrated components of the polarization, displayed in Eq.~(\ref{eq:ptotfi}) get corrections of around $0.1\%-0.2\%$. In all cases, the FBR correction is practically imperceptible.

To close this paper, it should be stressed that Eq.~(\ref{eq:ptot}) is general enough to be quite useful for processes where the momentum transfer is not small and thus cannot be neglected. Therefore it is valid not only for the semileptonic decay of $K^+$, but also of heavier mesons such as $D^+$ or even $B^+$, where it could provide a very good first approximation. To first order in $q$ it yields terms of order $\mathcal{O}(\alpha/\pi)(q/M_1)]$ in the RC. The expected error by the omission of higher order terms is around $(\alpha/\pi)(q/M_1)^2\approx 0.0012$ for $K_{\mu 3}^+$ decays. Being conservative, if the accompanying factors amount to 1 order of magnitude increase, then the upper bound to the theoretical uncertainty of 1.2\% can be estimated. This should be acceptable with an experimental precision of $2\%$-$3\%$.

\begin{acknowledgments}
The authors are grateful to Consejo Nacional de Ciencia y Tecnolog{\'\i}a (Mexico) for partial support. M.J.S.-G., J.J.T., and A.M.\ were partially supported by Comisi\'on de Operaci\'on y Fomento de Actividades Acad\'emicas (Instituto
Polit\'ecnico Nacional). R.F.-M.\ was also partially supported by Fondo de Apoyo a la Investigaci\'on (Universidad Aut\'onoma de San Luis Potos{\'\i}).
\end{acknowledgments}

\appendix

\section{\label{sec:in}The function $\theta_{24}$}
\subsection{The function $\theta_{24}$ in the TBR}

The function $\theta_{24}$, introduced in Eq.~(\ref{eq:AA}) arises from a new integral $I_N$ coming from the term $\omega^2/(1-\beta x)^2D$ appearing in the first summad of Eq.~(\ref{AL}). Explicitly, the integral reads
\begin{equation}
I_N = \int_{-1}^1 \frac{dx}{(1-\beta x)^2} \int_{-1}^{y_0} F^2 dy \int_0^{2 \pi} \frac{ d\phi_k}{D^3}.
\end{equation}
Following the procedure of Ref~\cite{tun89} is possible obtain
\begin{equation}
I_N = 2\pi \left[ 2(3E^2-l^2) \theta_2 - 6E^2 (2\theta_3 - \theta_4) + \theta_{24} \right],
\end{equation}
where $\theta_{24}$ is a new function that completes the ones defined in Ref.~\cite{tun89,tun91,tun93,torres04}; it reads
\begin{equation}
\theta_{24} = \int_{-1}^{1}dx \frac{\xi_3(x)}{(1-\beta x)^2},
\end{equation}
with $\xi_3(x)$ defined in Ref.~\cite{tun89}.

The analytical form of $\theta_{24}$ is
\begin{eqnarray}
\theta_{24} & = & \frac{6E_\nu^0l}{p_2}\left[(1+a^-)\frac{E}{E_\nu^0}(I_4-I_1)-2I_{2-}-(x_0+a^-)I_4+\frac{2(lx_0+E_\nu^0)}{l}I_{1-} + \frac{E_\nu^0}{3l}\left[-(x_0+a^+)^2I_{2-}^++(x_0+a^-)^2I_{2+}^-\right] \right] \nonumber \\
&  & \mbox{} - \frac{2l^2}{p_2} \left\{ a^+[(a^+)^2-1]I_{1-}^+ -a^-[(a^-)^2-1]I_{1+}^-\right\}, \label{eq:th24}
\end{eqnarray}
$x_0$, $a^\pm$, $I_1$ and $I_4$ are listed in Appendix B of Ref.~\cite{martinez00}. The other functions involved in $\theta_{24}$, Eq.~(\ref{eq:th24}), are

\begin{eqnarray}
I_{n-}&= & \int_{x_0}^1 \frac{x^{n-1}dx}{(1-\beta x)^2} \nonumber \\
& = & \frac{1}{\beta^{n-1}} \left[\frac{ (1-x_0)}{(1-\beta)(1- \beta x_0)} + \frac{n-1}{\beta}\ln \left|\frac{1-\beta}{1-\beta x_0}\right|\right],
\end{eqnarray}
for $n=1,2$.
\begin{eqnarray}
I_{1-}^+ & = & \int_{x_0}^1 \frac{dx}{(1-\beta x)^2(x+a^+)} \nonumber \\
& = & \frac{1}{(1+a^+ \beta)^2} \left[ \frac{\beta(1+a^+\beta)(1-x_0)}{(1-\beta)(1-\beta x_0)} + \ln\left|\frac{(a^+ +1)(1-\beta x_0)}{(1-\beta)(a^+ +x_0)}\right|\right],
\end{eqnarray}
\begin{eqnarray}
I_{1+}^- & = & \int_{-1}^{x_0} \frac{dx}{(1-\beta x)^2(x+a^-)} \nonumber \\
& = & \frac{1}{(1+a^- \beta)^2} \left[ \frac{\beta (1+a^-\beta)(1+x_0)}{(1+\beta)(1-\beta x_0)} + \ln\left|\frac{(a^- +x_0)(1+\beta )}{(1-\beta x_0)(a^- -1)}\right|\right],
\end{eqnarray}
\begin{eqnarray}
I_{2-}^+ & = & \int_{x_0}^1 \frac{dx}{(1-\beta x)^2(x+a^+)^2} \nonumber \\
& = & \frac{1-x_0}{(1+\beta a^+ )^2} \left[ \frac{1}{(a^++1)(x_0+a^+)}+\frac{\beta^2}{(1-\beta)(1-\beta x_0)} \right] + \frac{2\beta}{(1+\beta a^+)^3} \mathrm{ln} \left[\frac{(a^++1)(1-\beta x_0)}{(x_0+a^+)(1-\beta)}\right].
\end{eqnarray}
and finally,
\begin{eqnarray}
I_{2+}^- & = & \int_{-1}^{x_0} \frac{dx}{(1-\beta x)^2(x+a^-)^2} \nonumber \\
& = & \frac{1+x_0}{(1+\beta a^-)^2} \left[ \frac{1}{(a^--1)(x_0+a^-)} + \frac{\beta^2}{(1+\beta)(1-\beta x_0)} \right] - \frac{2\beta}{(1+\beta a^-)^3} \mathrm{ln} \left[ \frac{(a^--1)(1-\beta x_0)}{(x_0+a^-)(1+\beta)} \right].
\end{eqnarray}

\subsection{\label{sec:inb}The function $\theta_{24}^F$ in the FBR}

The function $I_N^F(E,E_2)$ is the FBR version of $I_N(E,E_2)$. In the integral form, $I_N^F(E,E_2)$ reads
\begin{equation}
I_N^F = \int_{-1}^1 \frac{dx}{(1-\beta x)^2} \int_{-1}^1 F^2 dy \int_0^{2 \pi} \frac{ d\phi_k}{D^3},
\end{equation}
whereas its analytical version becomes 
\begin{equation}
I_N^F = 2\pi \left[2(3E^2-l^2) \theta_2^F - 6E^2 (2\theta_3^F - \theta_4^F) + \theta_{24}^F \right], 
\end{equation}
where
\begin{equation}
\theta_{24}^F = \int_{-1}^1 dx \frac{\xi_3^F(x)}{(1-\beta x)^2},
\end{equation}
with $\xi_3^F(x)$ given in Eq.~(A6) of Ref.~\cite{juarez96}.

The analytical result for $\theta_{24}^F$ can be organized as
\begin{equation}
\theta_{24}^F = {T_{24}^+}^F + {T_{24}^-}^F,
\end{equation}
 with
\begin{equation}
{T_{24}^\pm}^F = 6(E_\nu^0-E) I_4 + 6E I_1 \mp 4 l a^\pm \frac{y_0^\pm}{{b^\pm}^2} [I_2^\pm+\beta I_1+\beta b^\pm I_4] \mp 2p_2\frac{{y_0^\pm}^2}{{b^\pm}^3} [2 \beta (I_2^\pm +\beta I_1)+b^\pm(I_3^\pm +\beta^2 I_4)].
\end{equation}
where the functions $a^\pm$, $b^\pm$, $y_0^\pm$, $I_1$, $I_2^\pm$, $I_3^\pm$ and $I_4$ are given in the Appendix B of Ref.~\cite{martinez00}.

\end{document}